\documentclass{aa}
\usepackage{graphicx}      
\usepackage{txfonts}
\usepackage{natbib}
\usepackage{epstopdf}
\usepackage{subfigure}
\usepackage{array}
\bibpunct{(}{)}{;}{a}{}{,} 

\title{Coupled Blind Signal Separation and Spectroscopic Database Fitting of the Mid Infrared PAH Features\thanks{This work is based on observations made with the Spitzer Space Telescope, which is operated by the Jet Propulsion Laboratory, California Institute of Technology under a contract with NASA.}}

\institute{Sterrewacht Leiden, Universiteit Leiden, Niels Bohrweg 2, NL-2333 CA Leiden, The Netherlands; \email{rosenberg@strw.leidenuniv.nl}\label{sterre}
\and 
The International Space University, Parc d'Innovation 1 rue Jean Dominique Cassini 67400 Illkirch Graffenstaden, France\label{isu}
\and 
NASA Ames Research Center, Space Science Division, Mail Stop 245-6, Moffett Field, CA 94035, USA; \email{Louis.J.Allamandola@nasa.gov, Christiaan.Boersma@nasa.gov}\label{nasa} }

\author{M.~J.~F.~{Rosenberg}\inst{\ref{sterre},\ref{isu}}
\and  O.~{Bern\'e}\inst{\ref{sterre}}
\and C.~{Boersma}\inst{\ref{nasa}}
\and L.~J.~{Allamandola}\inst{\ref{nasa}}
\and A.~G.~G.~M~{Tielens}\inst{\ref{sterre}}}

\date{Received:16-12-2010 /Accepted: 28-07-2011}

\begin{document}

\abstract{The aromatic infrared bands (AIBs) observed in the mid infrared spectrum of galactic and extragalactic sources are attributed to Polycyclic Aromatic Hydrocarbons (PAHs).  Recently, two new approaches have been developed to analyze the variations of AIBs in terms of chemical evolution of PAH species: Blind Signal Separation (BSS) and the NASA Ames PAH IR Spectroscopic Database fitting tool.}
{We aim to study AIBs in a Photo-Dissociation Region (PDR) since in these regions, as the radiation environment changes, the evolution of AIBs are observed.}
{We observe the NGC 7023-North West (NW) PDR in the mid-infrared (10 - 19.5 $\mu$m) using the Infrared Spectrometer (IRS), on board \emph{Spitzer}, in the high-resolution, short wavelength mode.  Clear variations are observed in the spectra, most notably the ratio of the 11.0 to 11.2 $\mu$m bands, the peak position of the 11.2 and 12.0 $\mu$m bands, and the degree of asymmetry of the 11.2 $\mu$m band.  The observed variations appear to change as a function of position within the PDR.  We aim to explain these variations by a change in the abundances of the emitting components of the PDR.  A Blind Signal Separation (BSS) method, i.e. a Non-Negative Matrix Factorization algorithm is applied to separate the observed spectrum into components.  Using the NASA Ames PAH IR Spectroscopic Database, these extracted signals are fit.  The observed signals alone were also fit using the database and these components are compared to the BSS components. }
{Three component signals were extracted from the observation using BSS. We attribute the three signals to ionized PAHs, neutral PAHs, and Very Small Grains (VSGs).  The fit of the BSS extracted spectra with the PAH database further confirms the attribution to PAH$^+$ and PAH$^0$ and provides confidence in both methods for producing reliable results.}
{The 11.0 $\mu$m feature is attributed to PAH$^+$ while the 11.2 $\mu$m band is attributed to PAH$^0$.  The VSG signal shows a characteristically asymmetric broad feature at 11.3 $\mu$m with an extended red wing.  By combining the NASA Ames PAH IR Spectroscopic Database fit with the BSS method, the independent results of each method can be confirmed and some limitations of each method are overcome.}


\titlerunning{Variations of 10-15 $\mu$m AIBs of \object{NGC 7023}} 
\authorrunning{Rosenberg et al.}

\maketitle

\section{Introduction}

Polycyclic Aromatic Hydrocarbons (PAHs) are carbonaceous macromolecules which were postulated to be present in the interstellar medium (ISM) in the 1980s (\citealt{Leger,  Allamandola, pug, alla}) and have since undergone an intense investigation in astronomy.  The state of the art and recent activity in the field of interstellar PAHs is well illustrated by the book ``PAHs and the Universe'', \citet{conf}.  Astronomical PAHs are generally considered to contain roughly 50 - 100 C atoms and have an abundance of a few $10^{-7}$ per H atom{\citep{TielensPAHreview}.}  Because of their nanometer size, the absorption of one far-ultraviolet (FUV) photon is sufficient to heat PAH molecules to high temperatures causing them to emit characteristic bands called Aromatic Infrared Bands (AIBs) which peak near  3.3, 6.2, 7.7, 8.6, and 11.2 $\mu$m (\citealt{TielensPAHreview} and references therein). PAHs are abundantly present in the diffuse ISM, reflection nebulae (RNe), planetary nebulae, protoplanetary disks, and galaxies. Observations of PAHs in Photo-Dissociation Regions (PDRs, \citealt{Hollenbach}), which are transition regions between atomic and molecular gas, but still strongly affected by the FUV photons, are of particular interest.  The UV flux decreases when moving from the neutral atomic gas to the dense molecular cloud and PAH populations will also evolve as the UV flux changes \citep{sloan, joblin}.  This effect is best studied in the mid-infrared (5 - 15 $\mu$m), where PAHs emit most strongly.  

Each AIB is charachteristic of a PAH vibrational mode (e.g. \citealt{pug, Allamandola, hony}), the 3.3, 8.6, and 10 - 15 $\mu$m features are due to the C-H stretching, in-plane and out-of-plane bending modes while the 6.2 and 7.7 $\mu$m features are mainly due to the C-C stretching modes.  These features have been observed to show strong variation in peak position, width of band (FWHM), and symmetry \citep{peeters3}.  The  changing ratio of the 8.6 and 11.3 $\mu$m features was discovered first and attributed to a change in the relative abundances of neutral and ionized PAHs \citep{joblin, sloan}.  Soon after, the 7.7/11.3 $\mu$m ratio was observed to vary as well, which was also attributed to the charge state of the PAHs.  Later, using \textit{Infrared Space Observatory (ISO)}, \citet{peeters3} catalogued the variations of the main features in the 6 - 9 $\mu$m range and empirically divided them into groups based on specific spectral properties.  It was found that each of these groups was representative of certain classes of objects: Class A included HII regions, RNe, galaxies, and non-isolated Herbig stars.  Class B included isolated Herbig stars, PNe, and two post AGB stars, while class C included only two post AGB stars.  Observations of the 10 - 15 $\mu$m region have been analyzed by \citet{hony} in terms of the solo, duo, trio, or quartet out-of-plane (OOP) bending mode of either PAH$^0$, PAH$^+$, or some combination of the two.  While the position and profile of these bands are quite characteristic the relative intensities do vary a lot, indicating variations in the edge structure of the aromatic molecules. 

Analysis and interpretation of astronomical observations is supported by dedicated laboratory studies and quantum chemical studies.  These studies are being carried out in many groups around the world, each using a different technique (see \cite{2011EAS....46...61O} for a review of laboratory and experimental studies).  In most cases, the absorption is studied at a low temperature in an inert matrix.  There are also some gas-phase experiments that have been carried out at higher temperatures.  These experimental studies have been extended by quantum chemical calculations using Density Functional Theory (DFT), to species not accessible in laboratory studies.  The DFT approach is used to determine the frequencies and intensities of vibrational modes.   Recently, these models have been used to calculate spectra for PAHs from 54 to 130 C atoms \citep{baus, baus09}.   This size range is particularly relevant for comparison to observations of space-based PAHs.   An extensive database has been created by the Astrophysics and Astrochemistry Laboratory at NASA Ames, which includes mid-IR to far-IR spectra of many different PAHs including large molecules, varied levels of ionization, and irregular shapes \citep{PAHDatabase}. \citet{2007CP....332..353M} used time dependent DFT methods and quantum-chemical calculations to report computed molecular properties of PAH emission for 40 molecules, available on an online database.  \citet{mulas} then modeled the PAH emission, which give positions and intensities of specific PAHs in different radiation environments.  The band profiles of some PAH emission were also calculated by \citet{2006A&A...456..161M}. 

Recently, the improved sensitivity of \textit{Spitzer Space Telescope} has brought a wealth of observations of AIB features.  NGC 7023 is a well studied and bright IR source where PAH variations are known to occur \citep{cesarsky}.  \cite{Werner} and \cite{sellgren} used \textit{Spitzer's Infrared Spectrograph (IRS)} in the Short-High (SH), Short-Low (SL), and High-Low (HL) modes to further observe the full mid-infrared range of \object{NGC 7023}.  They observed all the classical AIBs above 5 $\mu$m in addition to finding new, weak emission features at 6.7, 10.1, 15.8, 17.4, and 19.0 $\mu$m.  \citet{Olivier} observed NGC 7023, along with three other PDRs, using \textit{Spitzer's IRS-SL}.  The spectra were analyzed using a class of methods called Blind Signal Separation (BSS), which identifies elementary spectra from spectral cubes.  Using BSS on spectra of NGC 7023-NW, three component signals were recognized, PAH cations, neutral PAHs, and a third carrier which \citet{Olivier} attributed to evaporating Very Small Grains (VSG), following earlier assignment by \citet{Rapacioli}.  Although it is yet unclear what the exact nature of VSGs are, it has been proposed that they could be PAH clusters \citep{Rapacioli}.  
 
This paper presents a study of the PDR NGC 7023-NW, which aims to put observational constraints on the origins of the profiles and variations of the 10 - 15 $\mu$m spectra.  This study will complement and restrict previous results from quantum chemical calculations, ISO spectroscopy observations, \textit{Spitzer} IRS observations, and PAH models.  We analyze high resolution data from \textit{Spitzer's IRS-SH} \citep{Werner}, at a resolution, $R=\lambda/\Delta\lambda=600$, using the NASA Ames PAH IR Spectroscopic Database and a BSS method to separate the emitting components of the PDR.  

After briefly outlining the observational methods in Section 2, the Blind Signal Separation method is described in Section 3, including the application of BSS to the data (Section 3.2).  Section 4 presents our main results and compares these with previous studies of the region.  Section 5 provides a comparison to the fit with the NASA Ames PAH IR Spectroscopic Database.  Next, in Section 6, we discuss the implications of our results and propose strong candidates to explain the spectral variations of the 10-15 $\mu$m region.  Section 7 discusses briefly the nature of the VSG carrier and section 8 gives our concluding remarks.         
        
\section{Observations}
\begin{figure}
\resizebox{\hsize}{!}{
\includegraphics{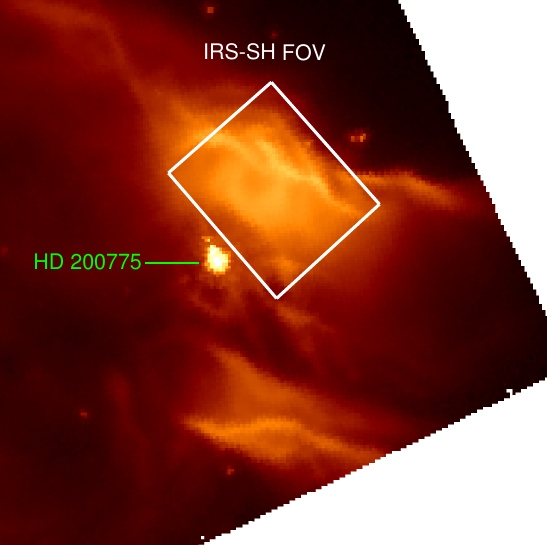}}
\caption{IRAC 8$\mu$m image of NGC 7023 \citep{Werner}.  Highlighted with a white box is the IRS-SH field of view around the PDR NGC 7023-NW.  The star, HD 200775, is marked by the green label.}
\label{fig:irs}
\end{figure}

NGC 7023-NW is a PDR located 40'' to the northwest of the exciting star, HD 200775, seen in Figure~\ref{fig:irs}.  HD 200775 is a magnitude 7 Herbig Be star and is located 430 pc from the sun \citep{1997A&A...324L..33V}.  There are 3 PDRs in NGC 7023 located east, south and northwest of the exciting star.  The northwest PDR is the brightest of the 3 PDRs.

NGC-7023-NW was chosen for this study in view of the interesting results of the analysis of low resolution data of this region performed by \citet{Olivier} who have shown that the AIB spectrum can be separated into 3 main components.  To further this study, high-resolution short wavelength (IRS=SH, 10 - 19.5 $\mu$m) observations were obtained from the Spitzer Heritage Archive (SHA).  The data was reduced using the CUbe Builder for IRS Spectra Maps (CUBISM), provided by \textit{NASA's Spitzer Space Telescope} tools \citep{smith}.  

\section{Methods}
\subsection{Blind Signal Separation}

Blind Signal Separation (BSS) is a class of methods used in many scientific fields to separate source signals from observed linear combinations of these signals e.g. separating brainwaves or unmixing recordings in acoustics. This has only recently been applied to astronomy \citep{firstbss, Olivier}.  In the case of observing a PDR, the resultant spectrum at any spatial point is a superposition of all elements emitting in the designated wavelength range and BSS can separate these components.  There are three main methods to perform BSS: Independent Component Analysis (ICA), Non-negative Matrix Factorization (NMF), and Sparse Component Analysis (SCA).  Based on the results of \citet{Olivier} and the added spectral resolution of the data, the NMF method was selected to perform the analysis.  If \textit{X} is a matrix containing the observed spectra, assuming that each spectrum is the result of a linear combination of source signals, we can write $X \approx WH$, where \textit{H} is the matrix of source signals, and \textit{W} is the matrix of mixing coefficients.  The goal of NMF is to recover $W$ and $H$ based on $X$ only.  The method we apply here is the same as \citet{Olivier} where all details can be found.

\subsection{Application to NGC 7023}

\begin{figure}
\resizebox{\hsize}{!}{
\includegraphics{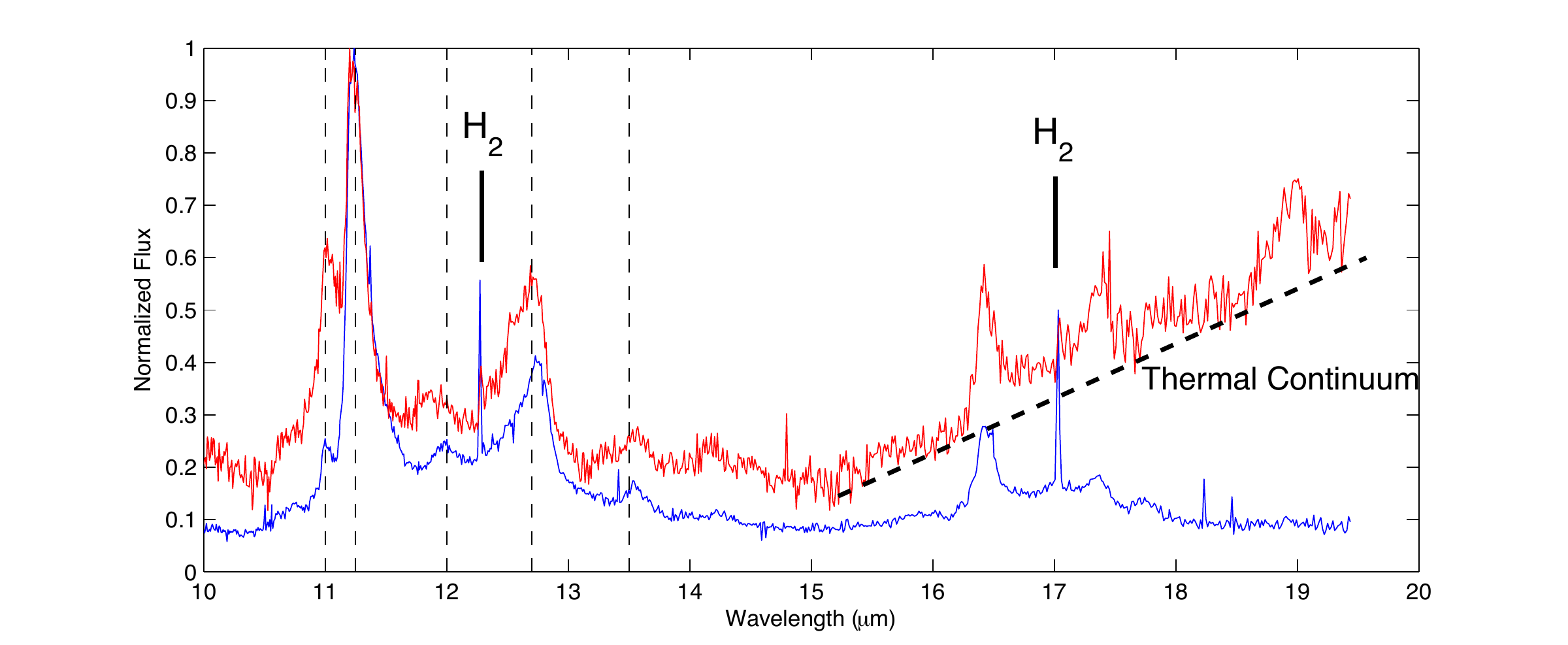}}
\caption{Observed spectra in pixel position [15,28] (blue) and [23,17] (red), containing H$_2$ lines and thermal continuum.}
\label{fig:spectra}
\end{figure}

Our astronomical data is comprised of 750 spectra, each taken at a different spatial location, and each spectrum including 869 wavelength points.  Figure~\ref{fig:spectra} shows two spectra, before they are decomposed with BSS.  There is a clearly defined separation between the 11.0 and 11.2 $\mu$m emission features as well as distinct features at 12.0, 12.7, 13.5, and 14.1 $\mu$m with additional features at longer wavelengths.  Among the features at longer wavelengths are the PAH 16.4 $\mu$m feature, the blended PAH and C$_{60}$ feature at 17.5 $\mu$m, and the pure C$_{60}$ feature at 18.9 $\mu$m, which are further discussed in \citet{2010ApJ...722L..54S}.  The relative intensities of the H$_2$ lines vary with position in the nebula, representing a non-linear component in our spectra.  Another non-linear aspect of the spectra is the onset of the dust grain continuum caused by heating from the source star.  These non-linear components cannot be analyzed by BSS methods, since the method demands a linear combination of signals.  Therefore, we have chosen to exclude the emission at wavelengths greater than 15 $\mu$m where the continuum from dust grains is present.  We have also clipped the H$_2$ lines everywhere in the cube by hand and replaced them by a linear interpolation.  The resulting spectra were then analyzed using the NMF algorithm from \citet{NMF} with both divergence and Euclidian distance optimization, see \citet{Olivier} for details. 

We investigated the possibility of 3, 4, 5, and 6 component source signals.  Irrespective of the particular minimization technique, when attempting to separate 4, 5 and 6 sources, there are 2 or more signals which are very similar (linear combinations of each other) and at least one signal that is pure noise.  Therefore, we can conclude that there are 3 significantly different spectral components responsible for the AIBs in NGC 7023-NW.  This result confirms the findings of \citet{Olivier} that there are only 3 source signals in this PDR.  

\section{Results of Blind Signal Separation}

\begin{figure*}
\centering
\includegraphics[width=17cm]{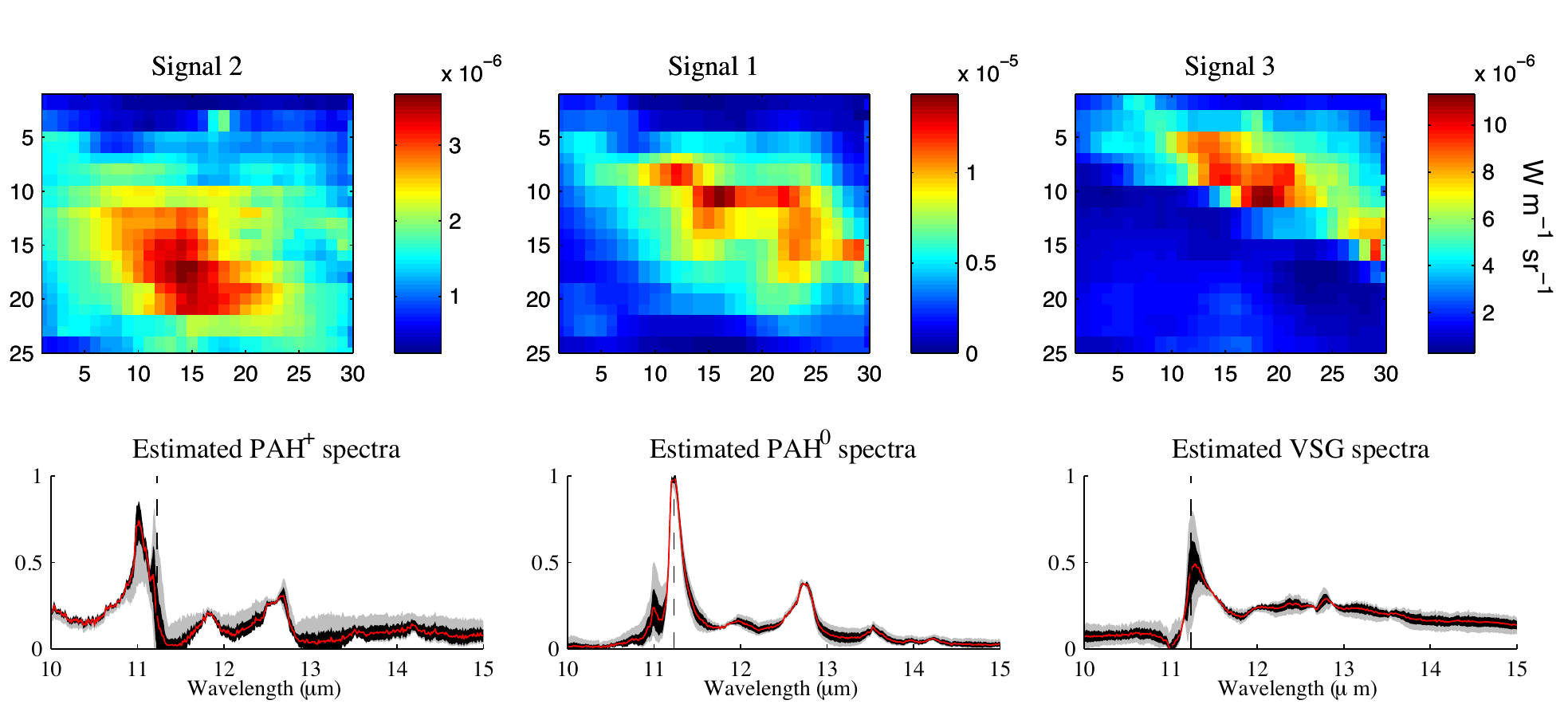}
\caption{\emph{Bottom Panels}: Extracted spectra using NMF, normalized at 11.2 $\mu$m.  The vertical line represents the peak position of the estimated PAH$^0$ spectra.  The red line represents the average spectra out of 100 iterations.  The grey envelope shows the minimum and maximum spectra and the black envelope shows the 1-$\sigma$ error of the 100 iterations.  \emph{Upper Panels}: Spatial distributions of the weighting factors obtained by Least Squares Fitting of the observed spectra in the datacube using the BSS extracted spectra shown in red in the lower panel.}
\label{fig:weightfactors}
\end{figure*}

\subsection{Extracted Source Signals}
\begin{figure}
\resizebox{\hsize}{!}{
\includegraphics{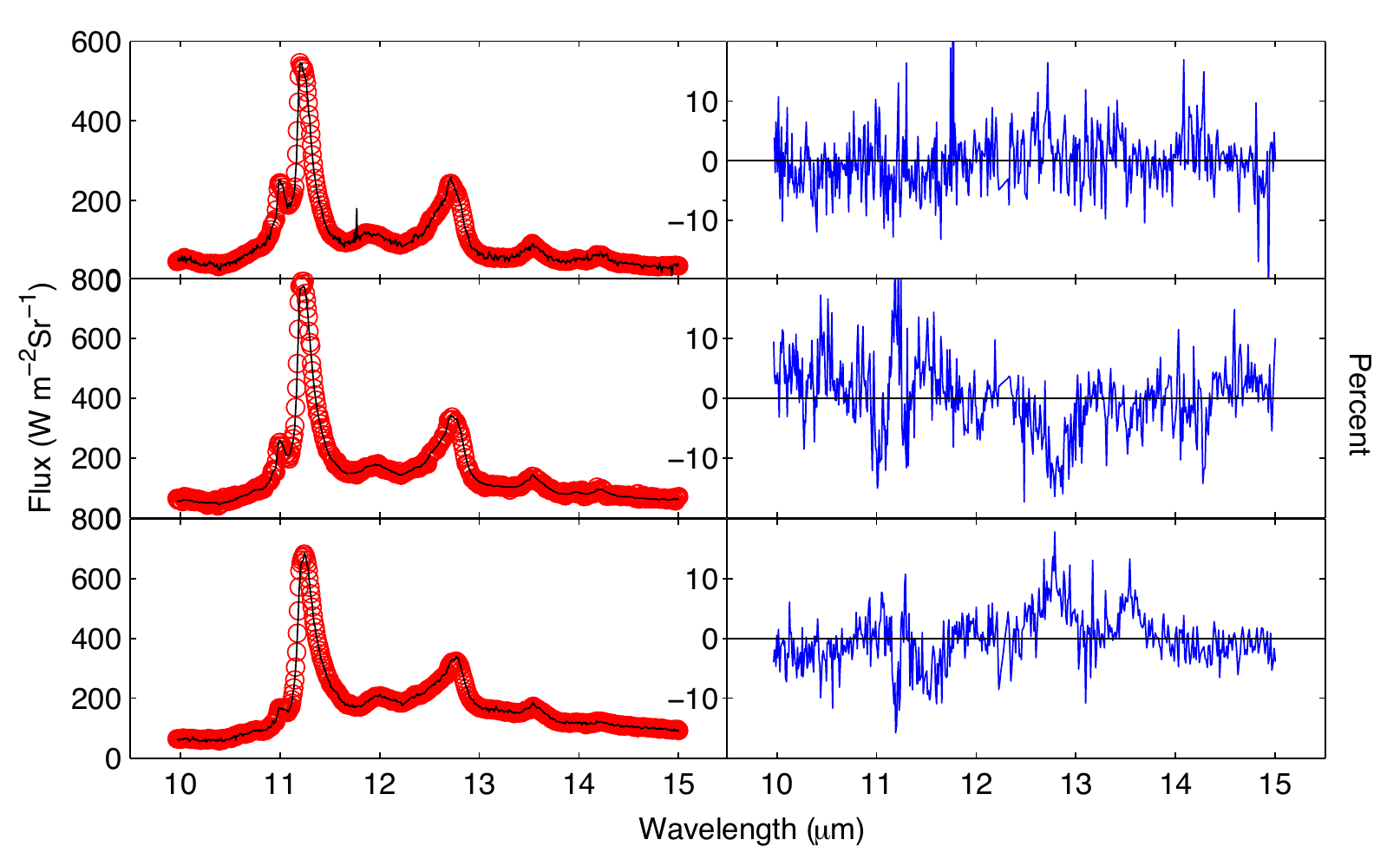}}
\caption{Spectra taken at three random spatial positions (left).  The solid line represents the original observed spectra while the red overlapping circles (thick red line) represent a linear combination of the BSS extracted spectra.  On the right is the matching residuals for each plot.}
\label{fig:residuals}
\end{figure}

The final extracted spectra are shown in Figure~\ref{fig:weightfactors}.  To increase confidence in these results, and ensure that this solution is not a random local minimum, the same analysis was repeated 100 times using different random initializations.  These 100 spectra shared the same general shape, but varied in intensity, especially in the 11.0 - 11.3 $\mu$m region.  The average spectra of the 100 iterations is plotted with a red line in Figure~\ref{fig:weightfactors}, and will be used for the remainder of our analysis as the final BSS extracted signals (H matrix).  We can also estimate the error at each point in the spectrum using these results (Figure~\ref{fig:weightfactors}).  The BSS method has the most difficulty separating the signals in the 11.0 to 11.3 $\mu$m range due to the strong changing spectral gradients there.  This results in large errors in this range (see Appendix~\ref{sec:appendixa} for discussion on unmixing artifacts).  Since $X \approx WH$, we can estimate $W$ by minimizing $\left \| X-WH \right \|$ using a standard least squares minimization.  Figure~\ref{fig:residuals} compares the observations in $X$ and the final reconstruction of these observations with $W \times H$.  The reconstruction is in good agreement with the observations.

Using the weighting factors that come as a resultant matrix of the above reconstruciton ($W$), we can map the spatial distribution of each source signal separately (Figure~\ref{fig:weightfactors}).  The spatial distribution shows clear variation for the three emitting components.   Signal 1 is most abundant in the middle of the PDR.  Signal 2 has its highest concentration closest to the source star (located at the bottom left of this image) and Signal 3 appears to trace the edge of the PDR farthest from the star.  The well defined regions where each signal is most concentrated implies a physical cause and gives further confidence that these results are not random.

\subsection{Carriers of the Extracted Spectra}
\begin{figure*}
\centering
\subfigure[]{
\includegraphics[width=5.66667 cm]{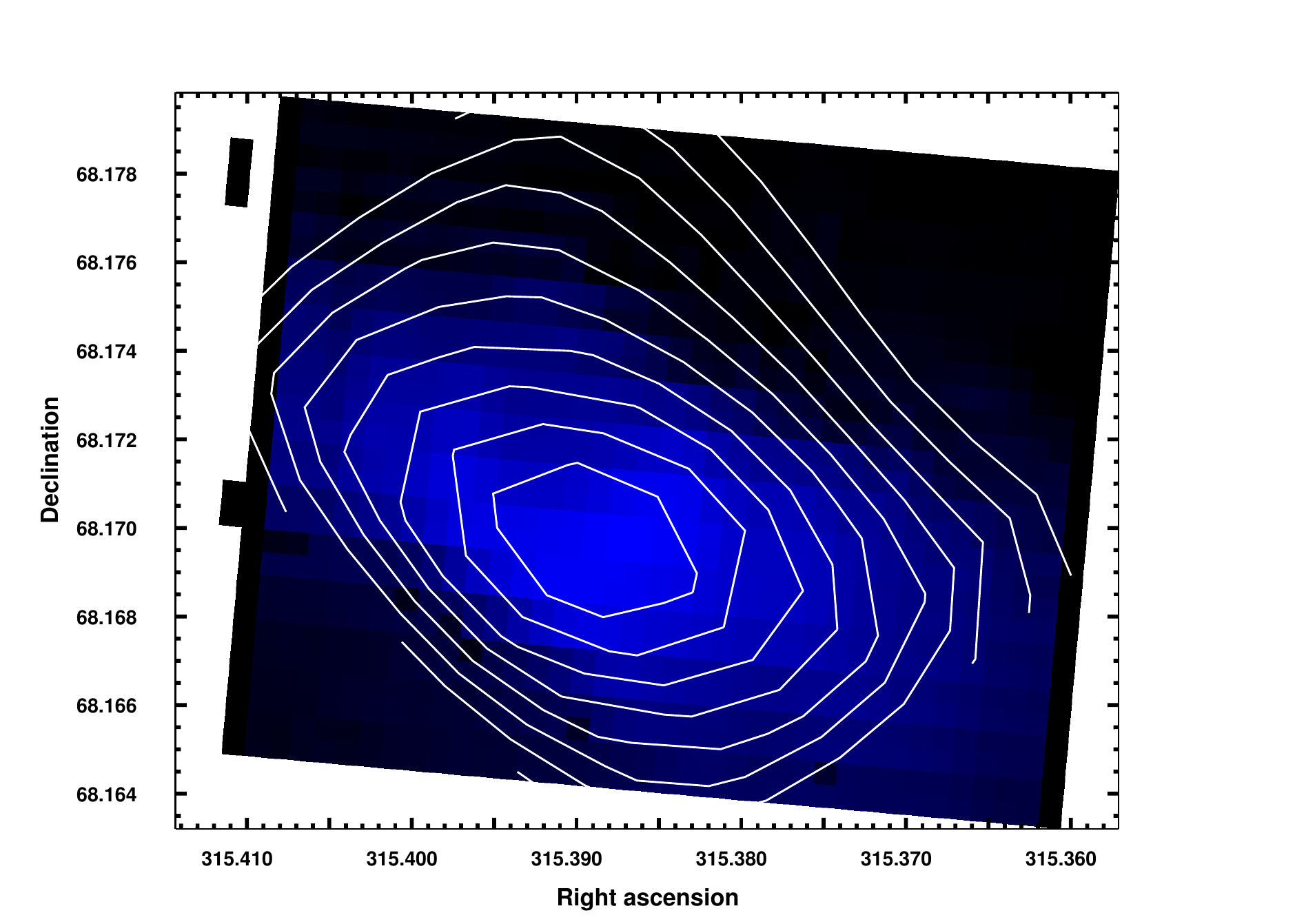}
\label{fig:subfig1}
}
\subfigure[]{
\includegraphics[width=5.66667 cm]{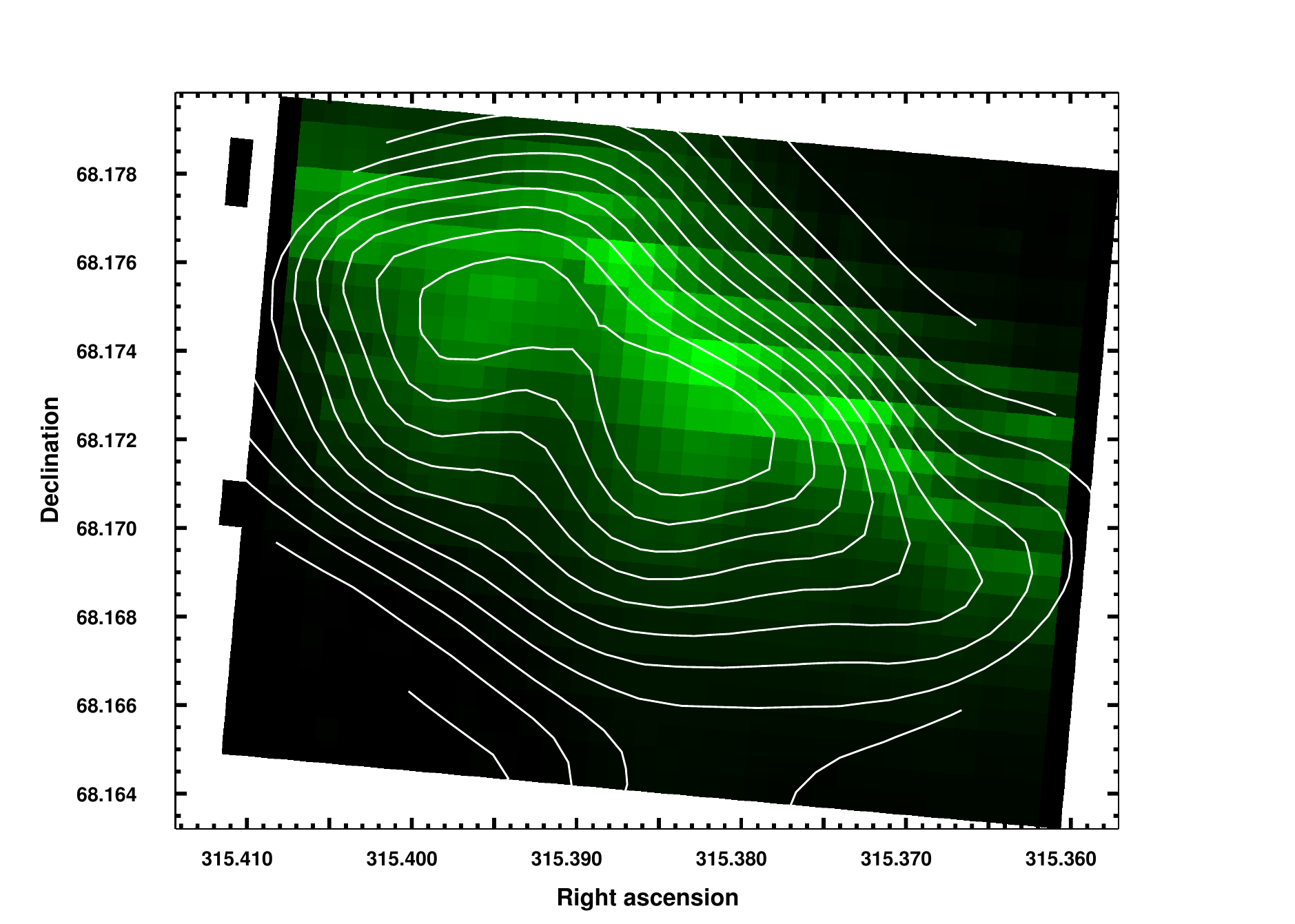}
\label{fig:subfig2}
}
\subfigure[]{
\includegraphics[width=5.66667 cm]{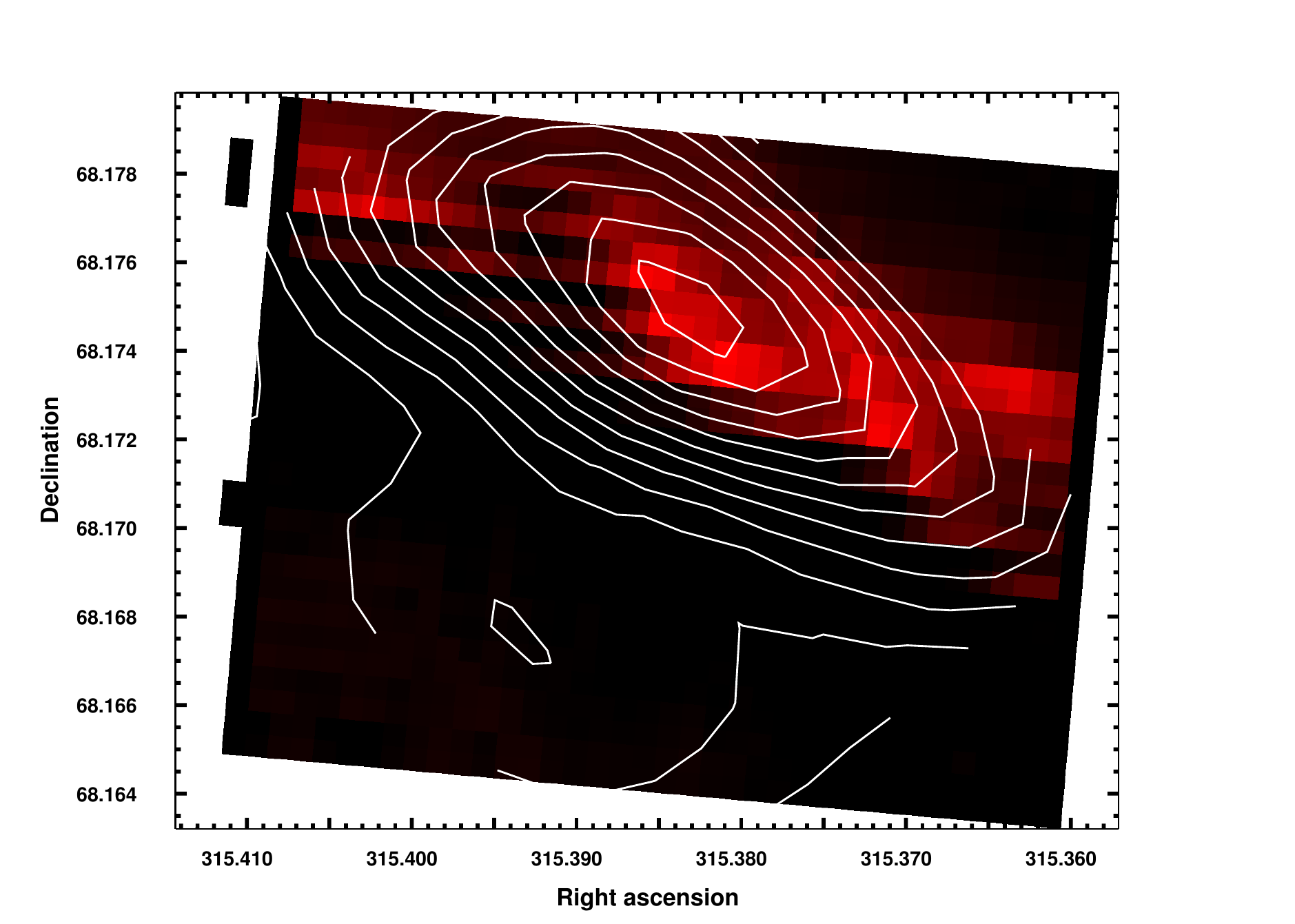}
\label{fig:subfig3}
}
\caption{(a) The contours of Signal 2, overlaid with the distribution of PAH cations created from \citet{Olivier}. (b) The contours of Signal 1, overlaid with the distribution of neutral PAHs created from \citet{Olivier}. (c) The contours of Signal 3, overlaid with the distribution of VSGs created from \citet{Olivier}.}
\label{fig:contours}
\end{figure*}

In this section, we will compare our results to the results of the low resolution study of the same region \citep{Olivier,Olivier10} to gain insight about the three extracted signals.  Creating spatial contours of intensity for each signal allows us to compare the spatial distribution of our signals to the distribution of the three signals from the study of \citet{Olivier} of the 5 - 15 $\mu$m low resolution spectra.  The contours are created from the IRS-SH spatial distribution maps (Figure~\ref{fig:weightfactors}) and overlaid with the spatial distributions (represented in color) of the IRS-SL results (Figure~\ref{fig:contours}).  The three signals extracted here show a strong spatial correlation to the PAH$^+$, PAH$^0$, and VSG maps of \citet{Olivier}.  The spatial distribution and the results of \citet{Olivier} seem to suggest that Signal 2 traces the distribution of PAH cations, Signal 1 the neutral PAH distribution, and Signal 3 the distribution of VSGs.  Although the spatial distributions of Signal 1, Signal 2, and Signal 3 correlate well with PAH$^0$, PAH$^+$, and VSGs of \citet{Olivier}, there are some small discrepancies, in particular, for the PAH$^0$ map. As discussed in \citet{Olivier10}, the degradation of spatial or spectral resolution always implies a loss in the quality of the NMF efficiency. Since \citet{Olivier} have a higher spatial resolution, while here we have a higher spectral resolution, none of the data-sets can be considered ``better'' and small discrepancies between the results of NMF are expected. 

\begin{figure*}
\centering
\includegraphics[width=17cm]{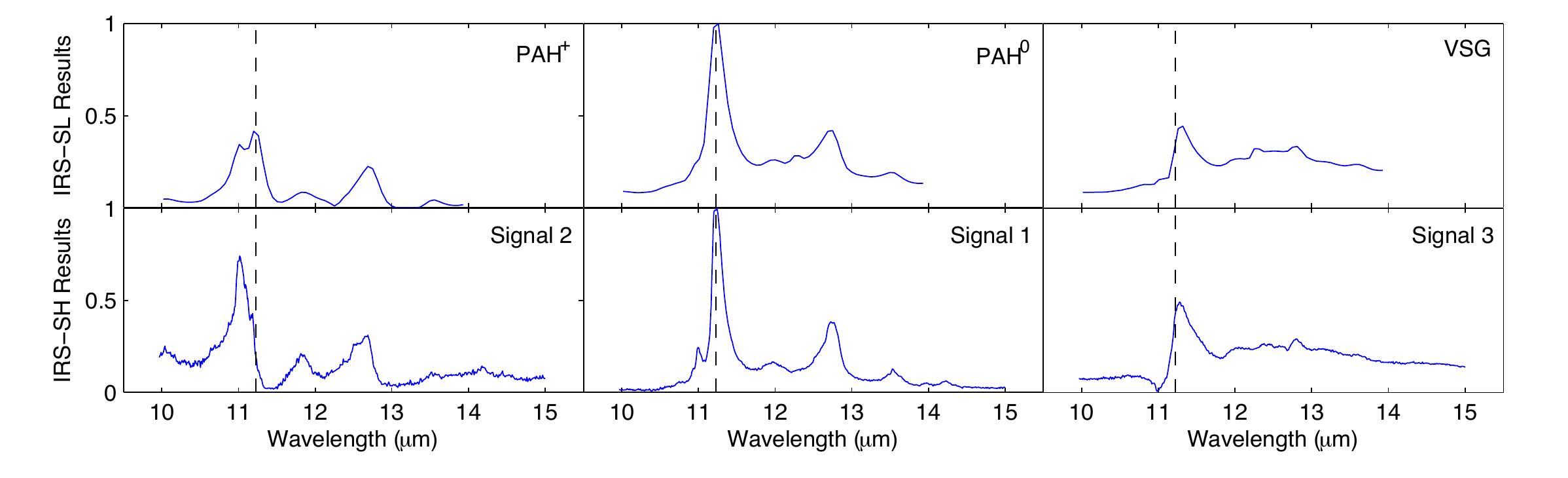}
\caption{The top spectra are the results from \citet{Olivier10} using low-resolution \textit{IRS} data, the bottom row of spectra are the current results, using high-resolution \textit{IRS}.  The vertical dashed black line indicates the 11.2 $\mu$m line position.}
\label{fig:spec_comp}
\end{figure*}

Figure~\ref{fig:spec_comp} compares the low-resolution source signal spectra of \citet{Olivier} to the high-resolution source spectra obtained here (Figure~\ref{fig:weightfactors}).  The low resolution extracted spectra share all the major features with the high resolution spectra, specifically the broad 11.3 $\mu$m emission feature in the VSGs, the 11.2 $\mu$m and 12.7 $\mu$m emission features in the neutral PAHs, and the 11.0 $\mu$m and broad 12.7 $\mu$m emission features in the PAH cations.  Although evidence for the 11.0 $\mu$m feature was present in the IRS-SL observations of \citet{Olivier10}, it was not immediately attributed to PAH cations since the resolution was not high enough to fully resolve and separate the 11.0 and 11.2 $\mu$m features.  The IRS-SH spectra has not only validated the previous results of the IRS-SL observations but given us additional spectral detail, to infer more about each signal and how it contributes to the observed spectra.  This is of particular interest in the 11 $\mu$m region where the 11.0 $\mu$m and 11.2 $\mu$m bands are well resolved and isolated.  We also observe a consistent broadening of the 11.3 feature in the VSG spectra.  One main difference in the spectra is the presence of an 11.0 $\mu$m satellite feature in Signal 1 that is not found in the IRS-SL PAH$^0$ signal, which we suggest is an artifact caused by the inherent limitation of unmixing and is compensated for by the sharp drop at 11.0 $\mu$m of Signal 3.  
The strong correlation between both the spectra and spatial distributions allow us to confidently identify Signal 1 as a PAH$^0$ signal, Signal 2 as a PAH$^+$, and Signal 3 as a genuine VSG spectrum.  This also matches the physical description of the PDR with the ionized species closest to the source star.       

\section{Comparison with Database Fitting Analysis}
\label{sec:dbfit}

The BSS-method allows the separation of the observed spectra into three mathematically distinct components, without taking the actual physical or spectroscopic properties of aromatic species into account. To further explore the carriers of the three signals resulting from the BSS analysis, we turned to the NASA Ames PAH IR Spectroscopic Database \citep{PAHDatabase}, which contains over 600 theoretical and experimental PAH spectra.  The existing extensive database of PAH spectra allows us to approach the analysis of the astronomical data from a different perspective.  Specifically, it allows us to link observational properties of the infrared emission features to the molecular characteristics of the carriers.  To that end, we first fit each BSS extracted spectrum with the PAH database.  Second, we focus on one observed spectrum from the IRS-SH data cube and use the database to fit this spectrum.  In interpreting the results from this database analysis, it should be kept in mind that vibrational modes in the mid-IR spectral range are characteristic for molecular groups and are not very sensitive to individual molecules.  Hence, the goal of this database analysis is to provide insights in trends rather than specific molecular identifications.  The trends of interest here involved the effects of charge, size, molecular geometry and symmetry -- including the degree of compactness of the PAH families -- on the different spectral components.  As a corollary, the completeness of the database is therefore of lesser concern, as long as the relevant classes are well represented.

The BSS method fits are ``blind'' in the sense that there is no a priori information about the nature of the signals built into the method. On the other hand, the fits of the database are based on spectra of actual aromatic molecules in specific charge states, structures, and sizes, allowing for a more direct interpretation of the results. 

The BSS results separated 3 mathematically distinct emitting signals and based on comparison with the signals extracted from \citet{Olivier}, we have attributed the three signals to PAH$^+$, PAH$^0$, and VSGs.  However, there is no aspect of the BSS method that actually identifies the signals as ionized or neutral PAH species.  By fitting these signals with the database, we can obtain an independent attribution of these distinct signals as PAH classes.

\subsection{Fit Parameters}
The database, at all versions, and the AmesPAHdbIDLSuite can be obtained from www.astrochem.org/pahdb. Here, version 1.11 of the theoretical component of the database and the November 10, 2010 version of the IDL suite was used. Briefly, the spectra in the theoretical database correspond to about 600 different PAHs, ranging in size from C$_9$H$_7$ to C$_{130}$H$_{28}$. Note that the database is biased towards smaller species, with PAHs containing over 50 carbon atoms making up roughly 24\%. The database includes PAHs at different charge states (i.e. cations, anions and neutral species) as well as different symmetries of the same molecule.  The fit included all C-H PAHs as well as polycyclic aromatic nitrogen heterocycles (PANH's; PAHs with one or more nitrogen atoms substituted into their carbon skeleton), since \citet{hudginsn} suggested that at least 1.2\% of the cosmic nitrogen is tied up in PAH molecules. The fit excludes PAHs with Oxygen, Magnesium, or Iron, where we note that no specific spectral evidence for the existence of such species has been found yet.  

Two approximations are made when fitting ``observed'' spectra with the database.  First, the spectra in the database refer to absorption spectra at 0 K, while the observed spectra are emitted by ``hot'' species. Due to anharmonicity, emission bands are observed to shift to the red with increasing temperature \citep{1992ApJ...401..269C,1998ApJ...493..793C}. Systematic experimental and quantum chemical studies on a very limited set of PAHs show that this redshift depends on the mode under consideration, the molecular structure and the temperature \citep{1995A&A...299..835J, 2003ApJ...591..968O,doi:10.1021/jp9088639}. The out-of-plane bending modes are observed to shift by about 15 to 20 cm$^{-1}$ for the small PAHs, pyrene and coronene between 0 and 900K and we will adopt this value for all PAHs in the database. Second, the database tools allow the user to specify a Gaussian or a Lorentzian profile with an assumed linewidth. However, the intrinsic line profile of PAH emission bands is distinctly non-Gaussian and non-Lorentzian due to the effects of anharmonicity and the accompanying red-shading \citep{1987ApJ...315L..61B, pech}. Here, we will adopt Lorentzian profiles, fully realizing that this implies that this procedure will force the fit of the observed broader and red-shaded bands to blends by emission from multiple species. We will adopt an intrinsic Lorentzian linewidth of 6 cm$^{-1}$ for the out-of-plane bending modes and note that this is somewhat less than the measurements ($\sim$10 cm$^{-1}$) at $\sim$700K for pyrene and coronene \citep{1995A&A...299..835J}. However, our ``choice'' to fit the observed, inherently asymmetric line profiles of the out-of-plane bending modes with symmetric calculated profiles forces us to adopt a somewhat small intrinsic line profile. We will assess the effect of these assumptions on our fitting results later on. Lastly, we mention that the relative strength of the bands in the calculated spectrum will also depend on the internal energy (eg., temperature) of the emitting species and hence on the absorbed photon energy and the temperature cascade. However, over the limited wavelength range considered here, this effect is very small and we will here simply adopt a single absorbed UV photon energy (6.5 eV) characteristic for the benign conditions of the PDR in NGC 7023 \citep{2009EAS....35..133J}. This corresponds to a peak temperature of $\sim$900K for a 50 C-atom PAH. The database evaluates the temperature for each PAH and follows the temperature cascade consistently.  Although our wavelength range is limited, we choose to use the temperature cascade to represented the most physically accurate approach. 

\begin{table}
\centering
\begin{tabular}{| l || c | c | c| r|}
\hline
Database & \multicolumn{3}{c|}{BSS Components} \\ \hline
 Percentage & PAH$^+$ & PAH$^0$ & VSG \\ \hline
 Large PAHs & 43 & 48 & 15 \\ \hline
 Small PAHs & 57 & 52 &  85    \\ \hline
 PAH Cations & 78 & 15 & 34 \\ \hline
 PAH Neutrals & 16 & 80 & 21\\ \hline
  PAH Anions & 6 & 5 & 45 \\ \hline
\end{tabular}
\caption{The numerical result of the fit using the PAH Database to fit the component spectra extracted using the BSS method.  The percentage amounts of large ($C\geq50$), small ($C<50$), cation, neutral, and anion species are displayed for the PAH$^+$, PAH$^0$, and VSG component signals.}
\label{tab:tab1}
\end{table}

\begin{figure*}
\centering
\includegraphics[width=17cm]{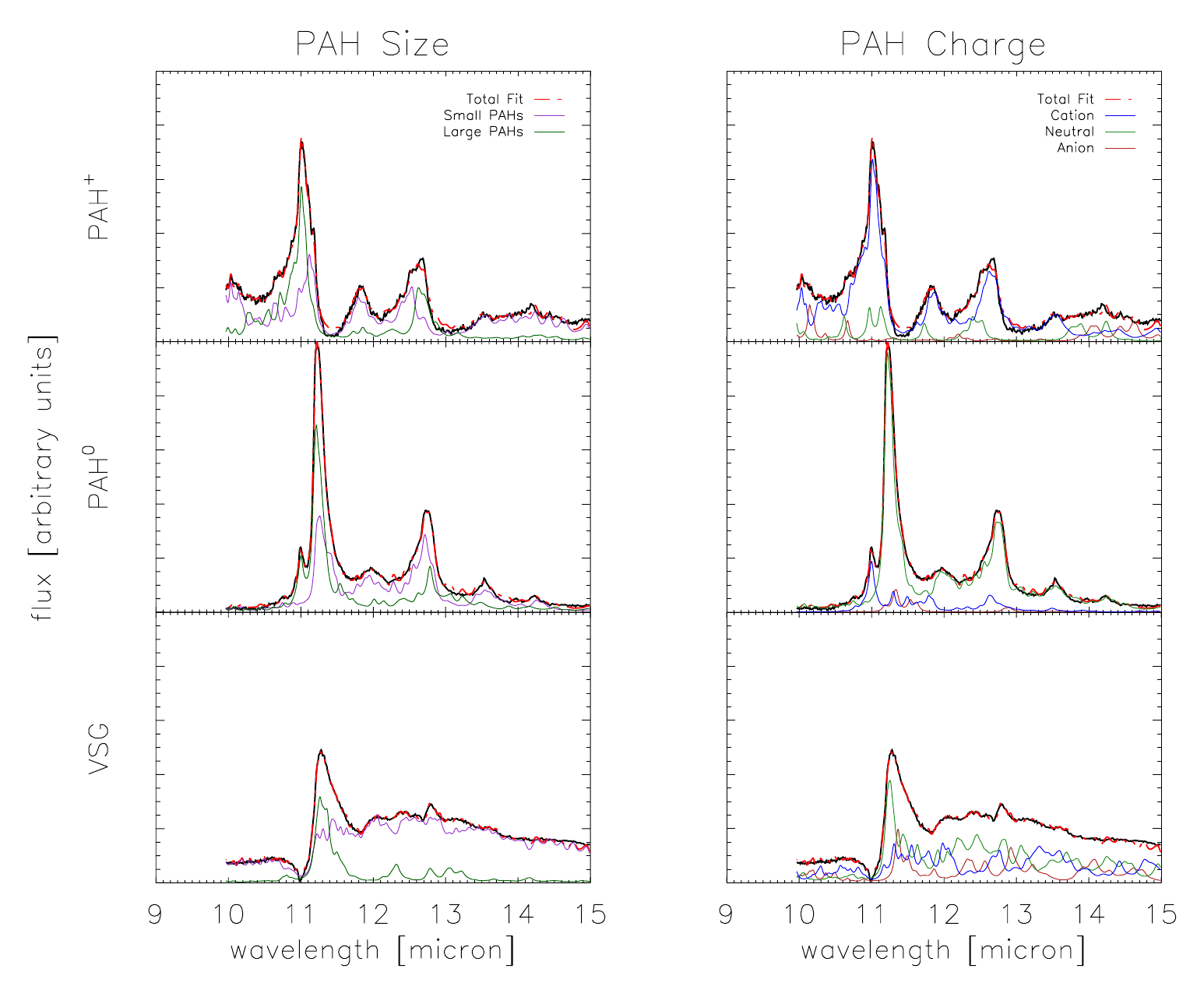}
\caption{The fit from NASA Ames PAH IR Spectroscopic Database \citep{PAHDatabase} of the three components extracted from BSS, PAH$^+$ (top), PAH$^0$ (middle), and VSG (bottom).  In the left column, we compare large PAH contribution to small PAH contribution.  In the right column we compare cation, neutral, and anion contribution to the fit.}
\label{fig:fit_matrix}
\end{figure*}

\subsection{Fit Results}
\begin{figure*}
\centering
\includegraphics[width=17cm]{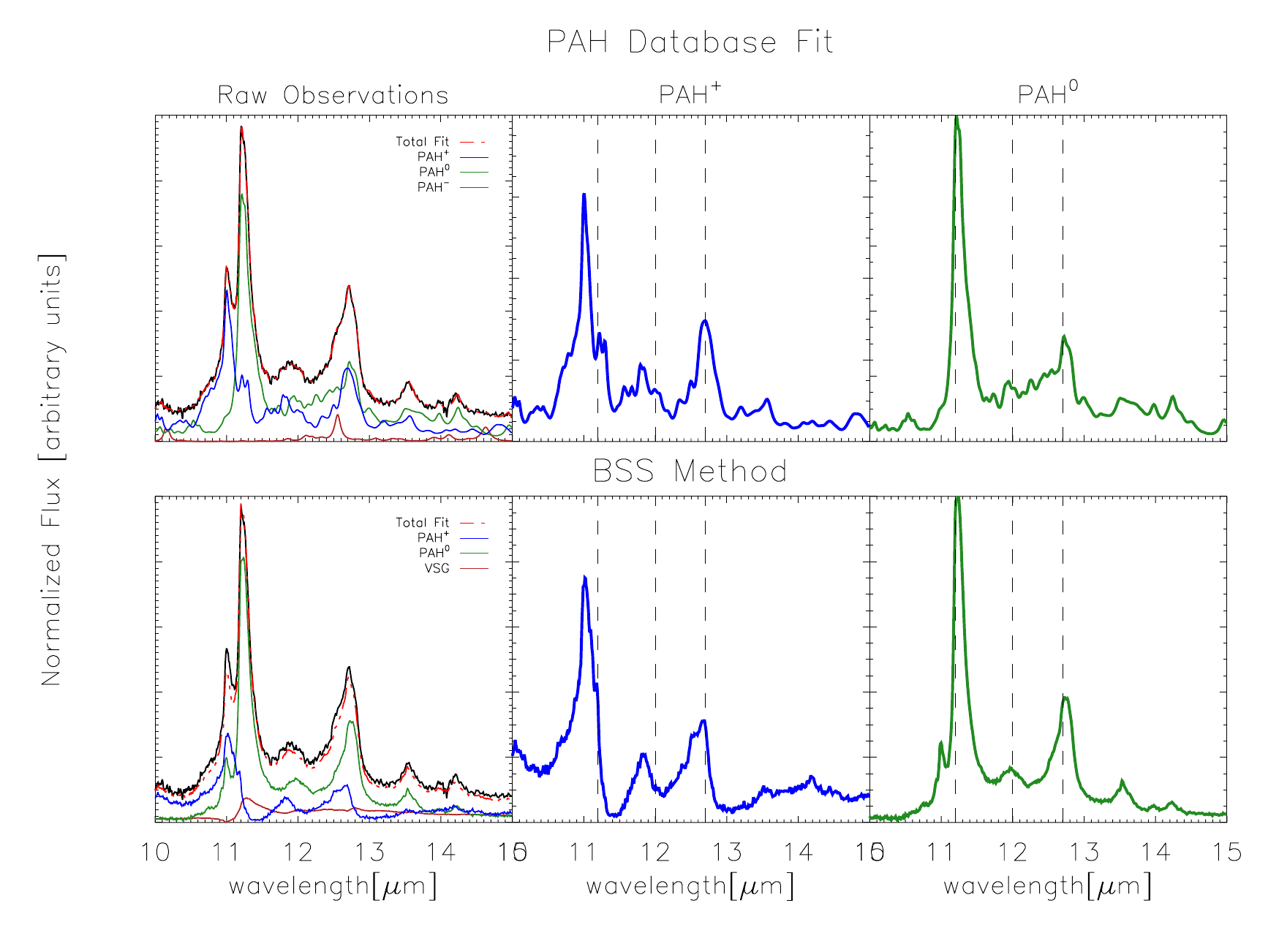}
\caption{Comparison of the fitting results of the two methods, the NASA Ames PAH IR Spectroscopic Database \citep{PAHDatabase} (top) and BSS (bottom).  The left panel depicts an observed spectra which is fit with both the database and the BSS extracted spectra.  The middle and right panels are the cation (middle) and neutral (right) components obtained from the respective methods.  The signals are normalized to the 11.2 $\mu$m peak.}
\label{fig:fitvbss}
\end{figure*}

 The results for the fits of the three extracted BSS component signals are presented in Figure~\ref{fig:fit_matrix}, broken down in categories of PAH size and charge. The small ( $<$50 C atoms) and the large ($\geq$50 C atoms) PAHs, as well as the cation, neutral, and anion species were separated out in order to judge each subgroupÕs contribution to the fit. A fit to the raw observed spectra at a typical position was also made (Figure~\ref{fig:fitvbss}). The results are summarized in Table~\ref{tab:tab1}. It should be emphasized that the contribution of individual species was quantified in terms of emission intensity, not in terms of abundance.  Figure~\ref{fig:fitvbss} shows a comparison between the PAH$^+$ and PAH$^0$ contributions of the database fit (top), and the PAH$^+$ and PAH$^0$ contributions in the BSS decomposition (bottom).  
The results of the database fits highlight various trends. First, except for the VSG spectrum, the overall contribution of large versus small PAHs seems very even. This may
arise from the three times greater number of smaller PAHs in the database. Remarkably,
although they represent only 25\% of the database, the main features of each spectrum
are generated by the larger PAHs while the continuum and less dominant features are
reproduced primarily by the smaller PAHs. Second, both the fit of the observed raw
spectra shown in Figure~\ref{fig:fitvbss} and the various PAH charge state contributions shown in the
right side of Figure~\ref{fig:fit_matrix} clearly shows that the dominant contributors to the 11.0 $\mu$m band are
PAH$^+$ while the bulk of the 11.2 $\mu$m feature is due to PAH$^0$ species. This behavior,
based on the spectra of hundreds of PAHs, confirms the early suggestion, based on a
handful of experimental PAH spectra, that the emission between 10.8 and 11.1 $\mu$m can be
used as a tracer of PAH cations \citep{hudgins}.  It should also be noted that the BSS extracted VSG
signal has two strong absorption features at about 11.0 and 12.6 $\mu$m. These probably
arise from cross mixing as described earlier and are likely artificial (Appendix \ref{sec:appendixa}). Therefore, the fit
was also performed with a linear interpolation over these points. This fit produced almost
identical results.

Common concerns regarding the PAH Database fitting methods are uniqueness and degeneracy.  In the case of PAHs with a limited wavelength range, these concerns must be approached in a different way.  This is further discussed and investigated in Appendix~\ref{sec:appendixb}.  These studies show that if the database is tasked to fit the BSS extracted PAH$^+$ spectra with only PAH$^0$ species, this is not possible, further strengthening the attribution of the 11.0 $\mu$m feature to PAH$^+$.   On the other hand, if the database attempts to fit the BSS extracted PAH$^0$ signal with only PAH$^+$, a suitable fit is provided.  However, without any limitations of the database fit, the reduced norm fit chooses predominately PAH$^+$ to fit the BSS extracted PAH$^+$ signal and PAH$^0$ to fit the BSS extracted PAH$^0$ signal (Table~\ref{tab:tab1}).  This only shows that PAH cations have emission features that peak through the 10 - 15 $\mu$m range and it is important to employ more than one method to apply as many astronomical constraints as possible.

\section{Origins of AIB Variations in the 10 - 15 $\mu$m Range}

Investigating the spectra and examining the spatial distribution of the extracted emission components, we recognize four aspects that vary with spatial position: the [11.0]/[11.2] $\mu$m ratio, the red wing of the 11.2 $\mu$m feature, the precise peak position of the 11.2 $\mu$m band, and the weak features at 12.0, 12.7, and 13.5 $\mu$m.  In the following section we discuss these variations in the context of Signal 1, 2, and 3, which will now be referred to as PAH$^0$, PAH$^+$, and VSGs, as emission feature carriers.  
\begin{figure}
\resizebox{\hsize}{!}{\includegraphics{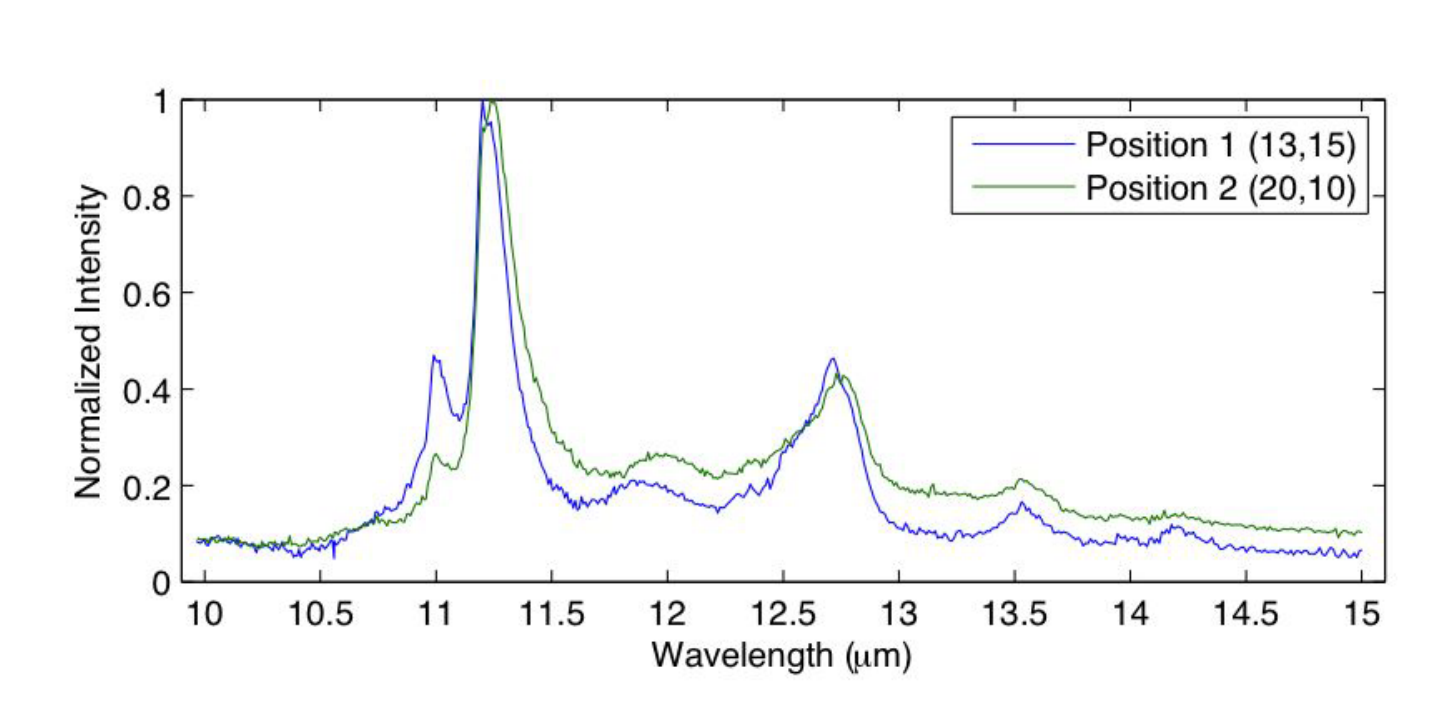}}
\caption{The original spectra from pixel positions (13, 15), where PAH$^+$ are highly concentrated and (20, 10), where VSGs are abundant highlighting the extremes of the observed spectral variations.}
\label{fig:analysis}
\end{figure}
\subsection{Ratio of 11.0 to 11.2 $\mu$m Features}
\label{sec:11.2}
In most observations of the 10 - 15 $\mu$m range, the dominant 11.2 $\mu$m feature appears with a dwarfed satellite feature at 11.0 $\mu$m.  This has been attributed to the solo C-H out-of-plane bending modes of PAH$^+$ \citep{hudgins,peeters, hony, baus, baus09}.  We observe the highest [11.0]/[11.2] $\mu$m ratio closest to the source star, which is also where the abundance of PAH$^+$ is greatest.  A comparison of observed spectra from a PAH$^+$ dominated region and a VSG dominated region are shown in Figure~\ref{fig:analysis}.  It is important to recall that regardless of position in the PDR, the 11.2 $\mu$m PAH$^0$ feature significantly dominates the 11.0 $\mu$m PAH$^+$ feature.  The separate contributions of PAH$^+$ and PAH$^0$ to the 11.0 and 11.2 $\mu$m features respectively, agrees with previous spectroscopic laboratory and quantum chemical calculation studies by e.g. \citet{hudgins}; \citet{hony}; \citet{peeters}; \citet{baus} and \citet{cami}.  In our decomposition analysis, the 11.0 $\mu$m band is clearly associated with the PAH$^+$ component (Signal 2) based upon the strong 6.2 and 7.7 $\mu$m bands, while the 11.2 $\mu$m band is attributed to the neutral component \citep{Olivier}.  Hence we suggest that the variation of the [11.0]/[11.2] $\mu$m ratio is due to a changing abundance of PAH$^+$ to PAH$^0$.

\subsection{The 11.2 $\mu$m Red Wing}
\label{sec:redwing}
As mentioned in the introduction, the 11.2 $\mu$m band has an observed asymmetry with a varying red wing \citep{roche}.  This wing has been attributed to anharmonicity or to different species of PAHs with a shifted solo mode peak emission \citep{pech, peeters}.  Observations of the 11.2 $\mu$m feature show that the shape and peak position can vary. In their analysis of the skewed variations in the 11.2 $\mu$m profile, \citet{peeters} empirically divided observations of the feature into two categories: one group is characterized by a peak between 11.20 and 11.24 $\mu$m and a more skewed red wing, while the other peaks at 11.25 $\mu$m and is much more symmetric.  In agreement with this, it has also been shown that PAH anions bands fall on the red side of the 11.2 $\mu$m peak and could contribute to the red wing \citep{baus}.   

By separating the observed spectra into component signals, we found that the main carrier of the 11.2 $\mu$m emission feature is PAH$^0$.  However, the observed spatial variations in the profile result in a PAH$^0$ source signal that is mainly symmetric with only a weak anharmonic red wing.  The BSS analysis shows that the red wing is mainly due to a changing contribution of the VSG to the observed spectra; e.g. as the VSG signal becomes more prominent towards the outer edge of the PDR, its contribution to the observed spectra also increases (c.f. Figure~\ref{fig:spec_comp}).  We note that in NGC 7023, position which are near to the exciting star are characterized by a more symmetric 11.2 $\mu$m profile while positions further away have a more pronounced red wing (cf., Figure~\ref{fig:analysis}).    

The feature at 11.2 $\mu$m has been observed to shift peak position between 11.2 to 11.3 $\mu$m.  Studying Figure~\ref{fig:spec_comp}, there are clear emission contributions from each component species throughout the 11.0 to 11.3 $\mu$m range.  We propose that the variation of abundances of VSGs and PAHs in the observed spectra causes the shifting peak position of the 11.2 $\mu$m band.  If there is a stronger contribution of PAHs in a certain region, the peak is observed to be blue-shifted.  If the VSGs become more abundant, the peak is redshifted.  This is in disagreement with \citet{peeters}, where they notice a redshifted peak with a more symmetric profile.

\subsection{The 12.0, 12.7, and 13.5 $\mu$m Features}
\citet{hony} assigned each emission feature to a different geometry and composition of PAH, depending on how many adjacent C-H groups are attached to the ring e.g., solo, duo, trio, and quartet modes of PAH$^0$ and PAH$^+$.  The results by \citet{hony} were further expanded to include compact and irregular shaped large PAHs by \citet{baus, baus09}, which are more astronomically relevant.  \citet{baus09} attributed the 11.3 - 12.3 $\mu$m band to the ``duo1" CH mode while the 12.5 - 13.2 $\mu$m region is attributed to ``duo2" CH$_{OOP}$ bands, the split of duo modes being caused by coupling to other bending modes.  The 13.5 $\mu$m feature has been attributed to the CH$_{OOP}$ quartet mode of large irregular PAHs \citep{baus09, hony} and can be used to place constraints on the edge structures of the emitting PAHs. Here we will place further astronomical constraints on the results of \cite{hony} and \cite{baus09}.

\subsubsection{The 12.0 $\mu$m Feature}  The peak position of the ``12.0 $\mu$m" feature varies from 11.8 $\mu$m to 12.0 $\mu$m (Figure~\ref{fig:analysis}).  The 12.0 $\mu$m feature has been attributed to both the PAH$^0$ and PAH$^+$ duo modes \citep{hony}.  Through the separation of source signals, we have isolated the main 12.0 $\mu$m feature to the PAH$^0$.  There is a feature that shares the profile of the 12.0 $\mu$m feature in the PAH$^+$ spectrum but it is blue-shifted, peaking around 11.8 $\mu$m.         

\subsubsection{The 12.7 $\mu$m Feature} The 12.7 $\mu$m feature has been predominately attributed to the overlap of PAH$^+$ duo and trio modes, but could not be definitely attributed to either PAH$^0$ or PAH$^+$ \citep{hony, baus, baus09}.  Examining the results of the signal separation in Figure~\ref{fig:spec_comp} and Figure~\ref{fig:fitvbss}, two unique features at 12.7 $\mu$m are revealed, that of PAH$^+$ and PAH$^0$.  Although they share roughly the same peak position, the PAH$^0$ 12.7 $\mu$m feature is shifted to the red and is seen from 12.5 to 13.0 $\mu$m, while the PAH$^+$ feature is blue shifted and asymmetric located between 12.3 and 12.8 $\mu$m.  We can attribute the variable blue wing of the 12.7 $\mu$m feature to the changing abundance of PAH$^+$.  A peak shift and prominent asymmetry is seen in Figure~\ref{fig:analysis} in Position 1, located in the most concentrated area of PAH$^+$.  This feature is seen along with the increased 11.0 $\mu$m feature, a blue-shifted peak position of the 11.2 $\mu$m feature, a decreased red wing of the 11.2 $\mu$m feature, and a ``12.0 $\mu$m'' feature peaking at 11.8 $\mu$m.  

\subsubsection{The 13.5 $\mu$m Feature} The observed spectra show a distinct 13.5 $\mu$m feature.  In a study of M17 it was suggested that this feature is coupled to the warm dust continuum \citep{cont}. \citet{hony} further investigated this possibility and instead, attributed the 13.5 $\mu$m feature to a quartet out-of-plane bending mode of PAH$^+$ and PAH$^0$.  Using BSS, we isolated this feature to PAH$^+$ (Signal 2) and PAH$^0$ (Signal 1), in agreement with the results of \citet{hony}, and likely decoupled from the warm dust continuum.   

\subsection{Systematic Blue Shift with Ionization}
We have attributed the 11.0 $\mu$m feature to PAH$^+$, while the 11.2 $\mu$m feature is attributed to PAH$^0$.  In addition, the 11.8 $\mu$m feature is attributed to PAH$^+$, while the 12.0 $\mu$m feature is attributed to PAH$^0$.  We also identify the broad 12.7 $\mu$m band in both PAH$^+$ and PAH$^0$, yet it appears to be a blend of features.  The PAH$^+$ 12.7 $\mu$m band is also bluer than the PAH$^0$ 12.7 $\mu$m band.  Specifically, the PAH$^+$ broad 12.7 $\mu$m feature spans 12.3 to 12.8 $\mu$m while the PAH$^0$ band stretches from 12.5 to 13.0 $\mu$m.  Comparing the PAH$^+$ and PAH$^0$ spectra, there is a systematic 0.2 $\mu$m blue shift between the emitting bands.  We do not observe this shift in the 13.5 $\mu$m band.  PAH band shifts can occur due to temperature change in the emitting region, yet according to the model proposed by \citet{pech}, a 0.2 $\mu$m shift of the 11.2 $\mu$m feature corresponds to a 650 K PAH temperature change.  This PAH temperature change is too great to be observed within NGC 7023 NW, therefore it is unlikely that this band shift is due to a temperature change.  Instead, we conclude that this shift is due to ionization, which modifies intrinsic emission properties of PAHs.  Investigation on the exact origin of this shift is, however, beyond the scope of our paper.

\subsection{Other Possible Effects on the Shape of AIBs in the 10 - 15 $\mu$m Range}
Other effects and chemical properties have been reported to alter the shape of AIBs in the 10 - 15 $\mu$m range.  Anharmonicity effects, as shown by e.g. \citet{pech} can modify the position and the symmetry of the 6.2 and 11.2 $\mu$m band and create the extended red wing in our observations.  By means of DFT calculations, [SiPAH]$^+$ $\pi$-complexes were also proposed by \citet{SiPAH} to produce a splitting of the initial 11.2 $\mu$m PAH band into two bands at 11.0 $\mu$m and 11.4 $\mu$m due to the Si adsorption on the PAH edge creates and a blue-shifted 6.2 $\mu$m band.  

We argue here (see Section~\ref{sec:redwing}), that the asymmetry of the 11.2 $\mu$m feature is predominately due to the contribution of VSGs.  Anharmonic effects are however still observed: the PAH$^0$ signal is not fully symmetric and displays a slight red wing, suggesting that anharmonicity effects are still important, but recall that most of the red wing is due to the varying abundance of the VSG component.  Since [SiPAH]$^+$ are expected to have a blue-shifted 6.2 $\mu$m band, we inspected the SL data but found no such signature.  The splitting of the 11.2 feature is seen in Signal 2 (PAH$^+$), which is most concentrated in the regions near the star.  Since the binding energy of [SiPAH]$^+$ is about 2 eV, they should be destroyed easily the highly irradiated environment near the star.  Altogether, this suggests that compact PAH$^+$ are a more natural explanation for the 11.0 $\mu$m feature, than [SiPAH]$^+$ $\pi$-complexes.

\subsection{Using the 11.0 and 11.2 $\mu$m features as tracers of ionization}
\begin{figure}
\resizebox{\hsize}{!}{\includegraphics[width=17cm]{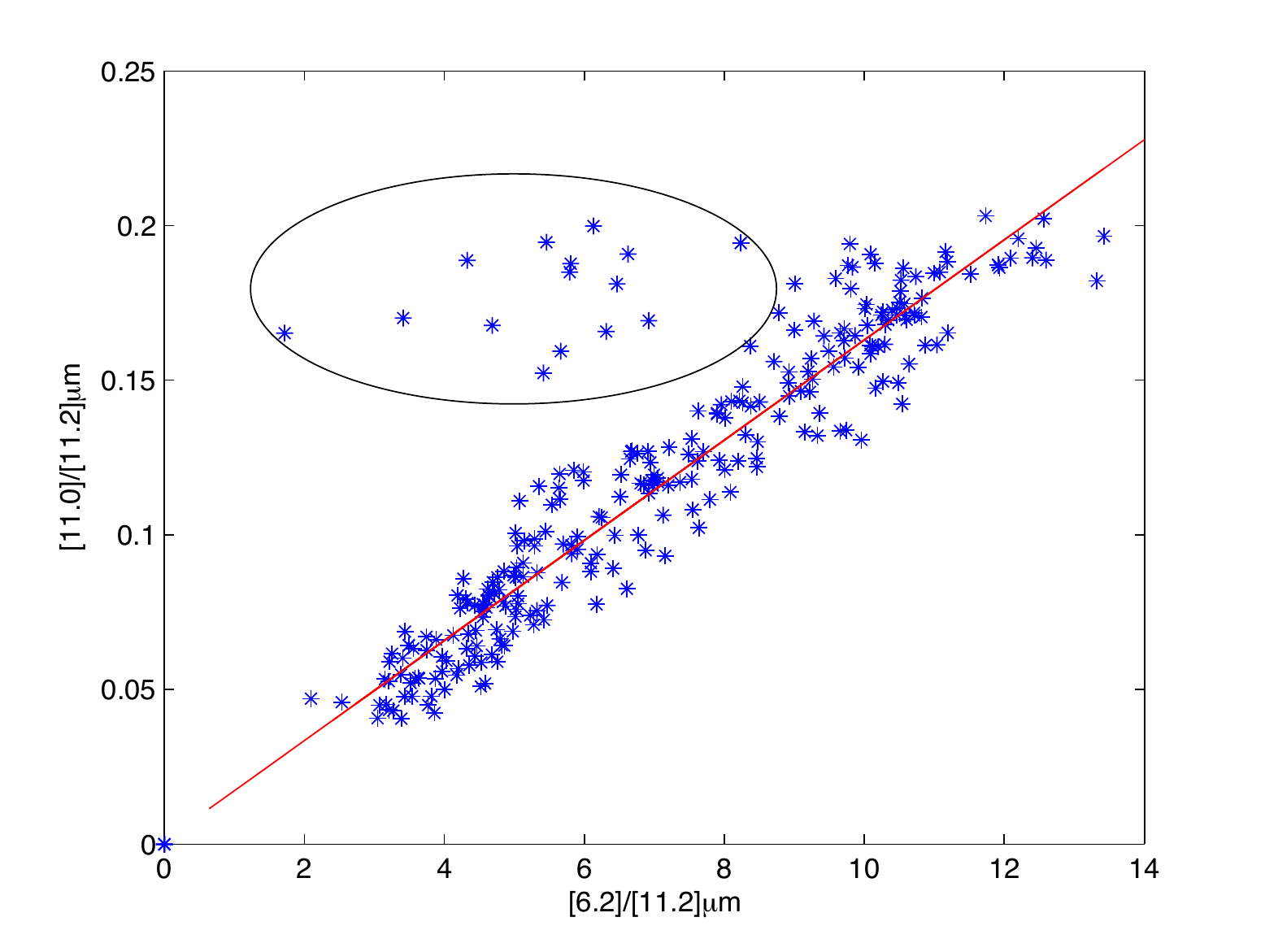}}
\caption{The [6.2]/[11.2] $\mu$m ratio vs the [11.0]/[11.2] $\mu$m ratio in NGC 7023-NW.  The 11.2 $\mu$m and 11.0 $\mu$m measurements were made using the IRS-SH observations while the 6.2 $\mu$m measurements were made using the IRS-SL observations.  The circled data were not included in the fit (see text for details).  The instrumental error is comparable to the symbol size.}
\label{fig:linear}
\end{figure}

With the attribution of the 11.0 $\mu$m feature to PAH$^+$ and the 11.2 $\mu$m to PAH$^0$, we can investigate the possibility of using this ratio to probe the ionization fraction of PAHs in the PDR.  One of the classic methods to trace the PAH ionization fraction is the [6.2]/[11.3] $\mu$m integrated intensity ratio (e.g. \cite{galliano}).  There are other tracers of ionization such as the [7.7]/[11.3] $\mu$m and [8.6]/[11.3] $\mu$m ratios, but the 6.2, 7.7, and 8.6 $\mu$m features include blended PAH$^+$ and PAH$^0$ bands.  As we show here, the 11.0 $\mu$m band is a purely cationic band and the 11.2 $\mu$m band is purely neutral, increasing the accuracy of ionization fraction measurements.  To demonstrate the reliability of the [11.0]/[11.2] $\mu$m ratio as an ionization indicator, we compare the [6.2]/[11.2] $\mu$m ratio to the [11.0]/[11.2] $\mu$m ratio using the  IRS-SL and IRS-SH observations of the NGC 7023-NW (Figure~\ref{fig:linear}).  In order to have an accurate measurement of the 11.2 $\mu$m feature, without contamination from the 11.0 $\mu$m satellite feature, we compare the integrated intensity of the 6.2 $\mu$m feature from IRS-SL observations to the intensity of the 11.2 $\mu$m feature using the high-resolution observations, since the 11.0 was not resolved and separated in the IRS-SL observations.  The maps were re-gridded using \textit{Montage} so that each point of the SH map corresponds to the same spatial position on the SL map.  Only the highest signal to noise data were used in this plot.  For the 6.2 $\mu$m low-resolution map, we set a band integrated intensity threshold of 10$^{-6}$ Wm$^{-2}$sr$^{-1}$.  For the 11.0 $\mu$m high-resolution map we set a threshold of 10$^{-7}$ Wm$^{-2}$sr$^{-1}$ and the 11.2 $\mu$m high-resolution map has a threshold of 10$^{-6}$ Wm$^{-2}$sr$^{-1}$.  The [6.2]/[11.2] vs [11.0]/[11.2] $\mu$m ratio in NGC 7023 is presented in Figure~\ref{fig:linear}. The data reveal a clear correlation, validating the use of the [11.0]/[11.2] $\mu$m ratio as a PAH ionization indicator.  The outliers in the upper left corner correlate to spectra where the thermal continuum from the source star is contaminating the linear continuum subtraction.  For this reason, these points were not included in the linear fit.  The linear fit has a high correlation coefficient of 0.95, from which an empirical relation can be derived:

\begin{equation}
\frac{[11.0 \mu m]}{[11.2\mu m]}=0.016\times\left ( \frac{[6.2\mu m]}{[11.2\mu m]} \right ) 
\end{equation}               

\section{Nature of the VSGs}
The BSS analysis identifies an independent broad component underneath the well-known 11.2 and 12.7 $\mu$m bands.  Earlier BSS studies over a wider wavelength range and the spatial distribution of Signal 3 (Figure~\ref{fig:contours}) assigns this component to emission by VSGs, proposed to be PAH clusters. Early observations of the 11.2 $\mu$m feature and its underlying emission support the suggestion that this broad underlying pedestal arises from a separate component \citep{1985MNRAS.215..425R}. The PAH Database analysis provides some further insight in the character of the carrier of this broad component. In this analysis, the 11-15 $\mu$m pedestal emission is due to a large number of individual components originating in a wide variety of molecular edge structures (solo's, duo's, and trio's), which together blend in an indistinct broad emission bump from 11 - 15 $\mu$m. For this blend, the analysis selects relatively small species from the database. However, that is a selection effect.  Small PAHs have, by necessity, a preponderance of corner structures. In contrast, calculations for large PAHs have focused (for obvious reasons) on regular structures with long straight edges and consequently strong 11.2 $\mu$m bands and weak bands at longer wavelengths. We surmise that large irregular PAHs would equally fit the bill. The VSG component has been assigned to clusters of PAHs based upon an interpretation of the observed spatial distribution and the physical properties of clusters \citep{Olivier, Rapacioli,2006A&A...460..519R}. However, the spectroscopic properties of PAH clusters are presently unknown.  While in general their spectra might be expected to resemble those of the constituent PAH molecules making up the cluster, we surmise that steric hindrance may affect the frequencies of the out-of-plane CH bending modes. We realize that there is a hidden issue here: the spectral differences in the 11-15 $\mu$m range -- the broad and indistinct band in the VSG component versus the very distinct 11.2 and 12.7 modes of the PAHs -- implies a more complicated evolutionary relationship between the VSGs and the PAHs than simple evaporation.

\section{Conclusion}
Applying a BSS method to observations from Spitzer's Infrared Spectrograph, Short-wavelength High-resolution mode, we uncovered 3 component signals in the PDR NGC 7023-NW.  We found that each signal is most abundant in different regions of the PDR, depending on the radiation environment.  We identified the three component signals as PAH cations, neutral PAHs, and VSGs.  As the observed spectra suggest, the neutral PAHs dominate every region of the PDR, but are most heavily concentrated in between the PAH cations and VSGs.  Both the spectra and spatial maps of each signal show high correlation to the results using SpitzerÕs IRS-SL mode \citep{Olivier, Olivier10}, allowing us to use these results to verify our conclusions.   

To further explore the origin of the three resolved signals, we employed the NASA Ames PAH IR Spectroscopic Database.  The fit shows that the component spectra resolved by BSS could be recreated by an appropriate combination of specific classes of PAH spectra from the database.  Then, we used a database to fit an observed spectrum and grouped the individual molecules into charge class, then compared the spectra of the combined charge classes to the BSS extracted PAH$^+$ and PAH$^0$ signals.  The components were found to be very similar to the BSS extracted PAH$^+$ and PAH$^0$.   

Specific spectral properties are found for each population:

\begin{itemize}
\item We have attributed the 11.0 and 11.2 $\mu$m bands to cations and neutral species respectively.
\item We conclude that the variation of the [11.2]/[11.0] $\mu$m ratio depends on the relative abundances of PAH cations to neutral PAHs. 
\item The extended red wing seen on the 11.2 $\mu$m feature is attributed to the increasing abundance of VSGs and the broad 11.3 $\mu$m feature that is characteristic of this component.
\item The changing peak position of the 11.2 $\mu$m feature can also be explained by varying contributions from PAHs (blue shift) and VSGs (red shift).
\item The 12.0 $\mu$m feature is attributed to neutral PAHs while the 11.8 $\mu$m feature is attributed to PAH cations, therefore, as the ionization mixture changes, the peak of this feature will shift accordingly.
\item Since the 13.5 $\mu$m feature is present in both PAH cations and neutral PAHs, but not existent in the VSG signal, where we see the continuum, we agree with \citet{hony} that the 13.5 $\mu$m signal is decoupled from the 15 $\mu$m continuum.
\end{itemize}

By using the BSS method and the PAH Database fit, we arrived at the above conclusions.  Each method has unique yet complementary strengths and weaknesses.  The BSS method is blind, i.e. has no intrinsic assumptions about the emitting components, however since the statistical properties of the emitting components are unknown, the unmixing is not perfectly efficient.  Additionally, the BSS method separates 3 mathematically distinct signals, but gives no intuition about the molecular properties of these signals.  The PAH Database allows for direct physical interpretation of the fit yet is biased towards smaller molecules and lacks spectral information for PAH clusters or other possible carriers of the VSG signal.  Although both methods suffer limitations, the strengths of one compensate for the weaknesses of the other.   Although the database fit may be degenerated in some cases, the interpretation of the $\chi^2$ database fit results, in terms of classes, is in agreement with the result of the BSS for PAH$^+$ and PAH$^0$.  The VSG spectrum can only be obtained by BSS, and the database fit of this spectrum provide additional information on the possible chemical nature of this component.  Both methods, i.e. BSS and Database Fitting, are powerful tools, but they \emph{must} be used with an understanding of their limitations (described in details in Appendices A and B). 

\begin{acknowledgements}
This work was conducted by M.~Rosenberg in part fulfillment of the M.Sc. Degree in Space Studies at the International Space University (ISU), Strasbourg, France.  The author also acknowledges M.Sc. scholarship support from ISU.
L.~J.~Allamandola acknowledges support from NASA's Astrobiology and Laboratory Astrophysics Program.
We acknowledge B.~Joalland for fruitful discussions on the spectral properties of SiPAH complexes.
This research made use of Montage, funded by the National Aeronautics and SPace Administration's Earth Science Technology Office, Computation Technologies Project, under Cooperative Agreement Number NCC5-626 between NASA and the California Institute of Technology.  Montage is maintained by the NASA/IPAC Infrared Science Archive.
C. Boersma acknowledges support by an appointment to the NASA Postdoctoral Program at the Ames Research Center, administered by Oak Ridge Associated Universities through a contract with NASA.
ERC grant: Studies of interstellar PAH at Leiden Observatory are supported through advanced-ERC grant 246976 from the European Research Council.
The authors also thank the referee for their time, comments, and suggestions.
\end{acknowledgements}

\appendix
\section{Exploring the Artifacts of BSS}
\label{sec:appendixa}
The efficiency of NMF is subject to two main limitations: 1) the possible non-unicity of solutions 2) the inaccuracy of the unmixing in the presence of noise.  These two problems are the subject of intensive theoretical research in the field of signal processing (see e.g. \citealt{donoho} and \citealt{hans}).  If the statistical properties of the source matrix are known, then one can determine whether NMF will function properly.  In a ``real life" case like here however, since the matrix is not known \emph{a priori}, we cannot verify these statistical properties.  We therefore rely on some empirical considerations.  Procuring the component spectra (Figure~\ref{fig:weightfactors}), the most prominent unmixing artifact is located at 11.0 $\mu$m and reveals itself with a sharp drop in Signal 3.  Some less discernible artifacts can be found at 11.2 $\mu$m as an added peak of Signal 2 (seen in the grey envelope of Figure~\ref{fig:weightfactors}) and at 12.7$\mu$m as a dip in intensity of Signal 3, roughly resembling an absorption feature.  

\begin{figure*}
\centering
\includegraphics[width=17cm]{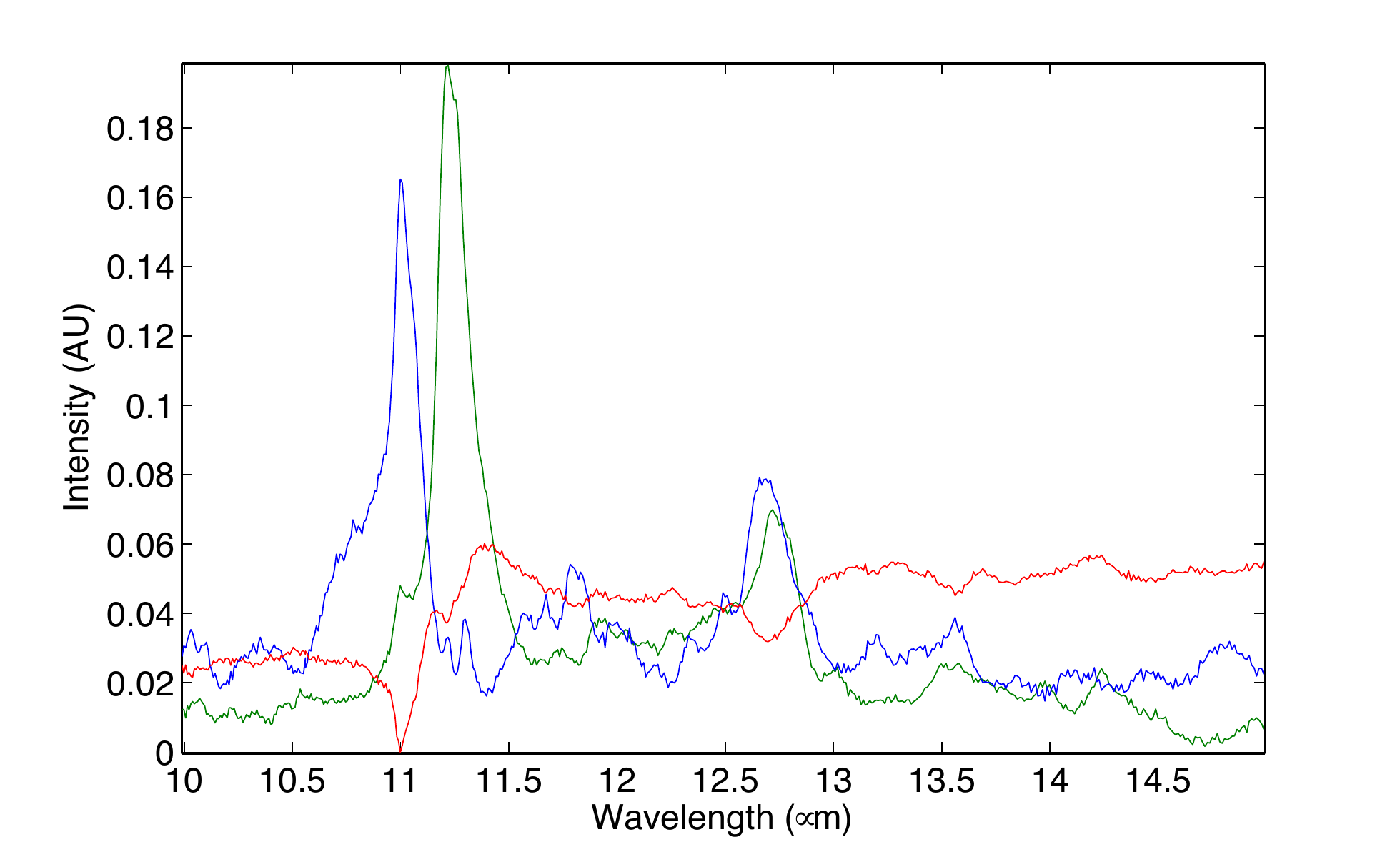}
\caption{The result of BSS using the NMF method on artificially mixed combinations of the PAH Database PAH$^+$ and PAH$^0$ signal and the VSG signal from \citet{Olivier}. BSS extracted PAH$+$ in blue, extracted PAH$^0$ in green, and extracted VSG in red.}
\label{fig:bss_proof}
\end{figure*}

\begin{figure*}
\centering
\includegraphics[width=17cm]{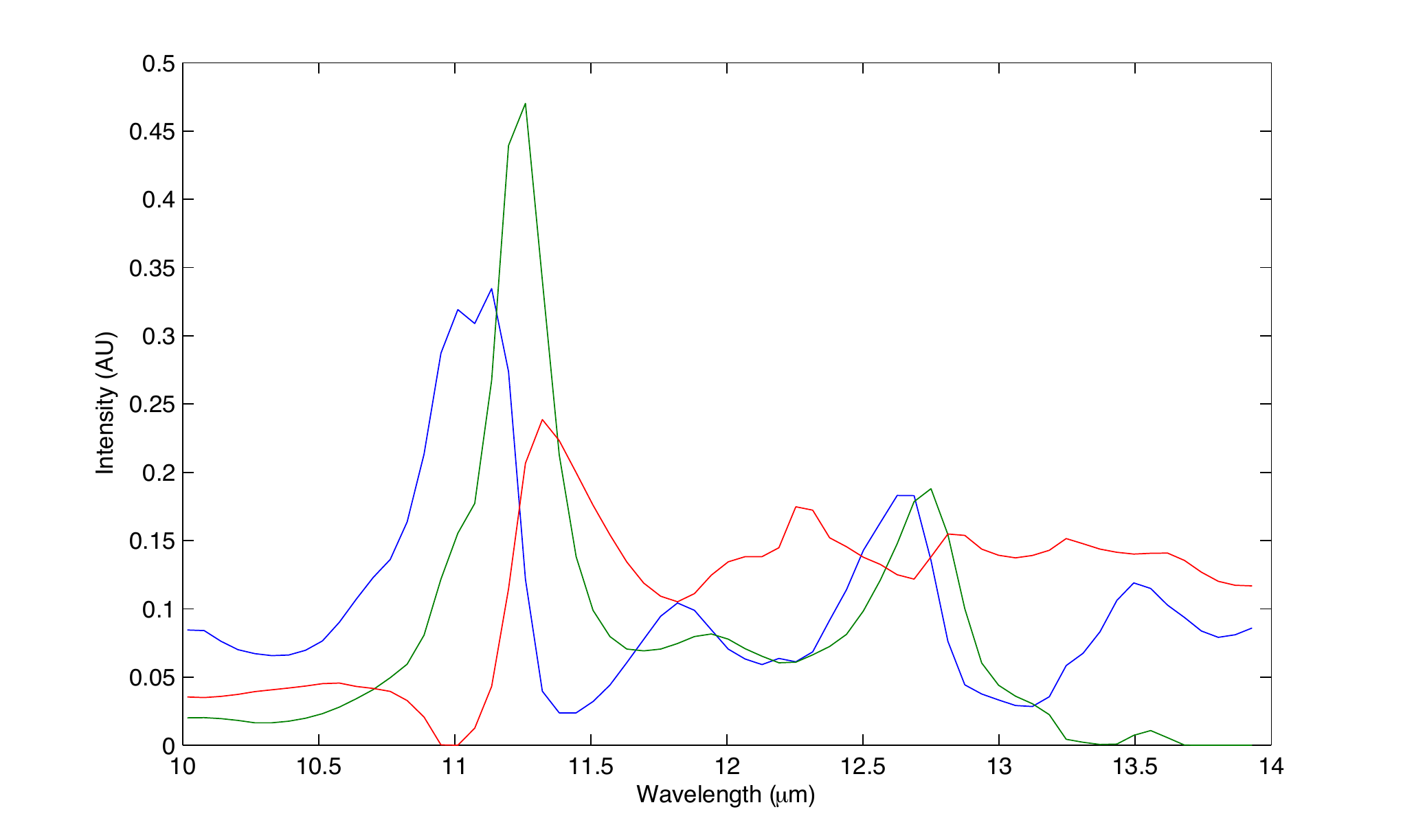}
\caption{The result of BSS using the NMF method on the IRS-SL data from \citet{Olivier} restricting the wavelength range to 10-14 $\mu$m.  BSS extracted PAH$+$ in blue, extracted PAH$^0$ in green, and extracted VSG in red.}
\label{fig:oliv_bss}
\end{figure*}

We suggest that the unmixing artifacts seen in some signals are compensated for by an increase or decrease of intensity, at the same wavelength, in other signals.  This is most clearly seen at the 11.0 $\mu$m wavelength.  The sharp drop to 0 of Signal 3 is compensated by the weak satellite feature of Signal 1.  Similarly, the absorption-type feature seen at 12.7 is compensated by a slightly increased intensity of Signal 2 at 12.7 $\mu$m.  In order to better understand the unmixing efficiency, two tests were conceived:  
\begin{enumerate}
\item We have artificially mixed the database PAH$^0$ and PAH$^+$ (Figure~\ref{fig:fitvbss}) signals with the
NMF extracted VSG signal of \citet{Olivier} (Figure~\ref{fig:spec_comp}).  We mix these signals with a $3\times100$ random matrix, but we make sure that
the coefficients correspond to realistic relative abundances (e.g. there are no regions with 100 \% PAH$^+$). We add synthetic white 
gaussian noise to the spectra to obtain a signal to noise ratio of about 20 (similar to our observations).  In this way, we created 100 simulated spectra. 
We then apply NMF to this set of spectra.  An example of the extracted signals is shown in Figure~\ref{fig:bss_proof}. There is a clear absorption band in the 
VSG spectrum, very similar to what is observed in the VSG spectrum extracted from the observations (Figure~\ref{fig:weightfactors}).  Lowering the noise does
not improve these results. This implies that this artifact is caused by improper unmixing, and that this is an inherent problem of the NMF
algorithm: the \emph{source} spectra we are trying to extract most likely do not verify completely the statistical properties that are needed
for NMF to function efficiently \citep{donoho}.  We conclude that the absorption feature we observe in the VSG extracted spectrum (Figure~\ref{fig:weightfactors}), is caused by the improper
unmixing of NMF. 
\item In order to test why \citet{Olivier} did not see such an artifact in their VSG spectrum, the same BSS NMF algorithm was used to separate the IRS-SL data cube components but with only the 10-14 $\mu$m wavelength range included.  The result of this BSS is shown in Figure~\ref{fig:oliv_bss}, where the 11.0 $\mu$m absorption artifact is clearly visible.  The fact that this artifact becomes present when only including the 10-14 $\mu$m spectral range implies a lack of information and constraints in this particular wavelength range.  \cite{Rapacioli} and \cite{Olivier} show that the main features of the VSG spectrum are 1) a broad 7.7 $\mu$m band, 2) a continuum, dominant from 15 to 20 $\mu$m.  Without these constraints in the reduced spectral range of IRS-SH, BSS has a difficulties extracting enough information for this signal.  We conclude that the artifacts especially in the VSG spectrum are mostly due to this reduced spectra range.
\end{enumerate}

\section{Exploring the Limitations of the PAH Database Fitting}
\label{sec:appendixb}

\begin{figure*}
\centering
\includegraphics[width=17cm]{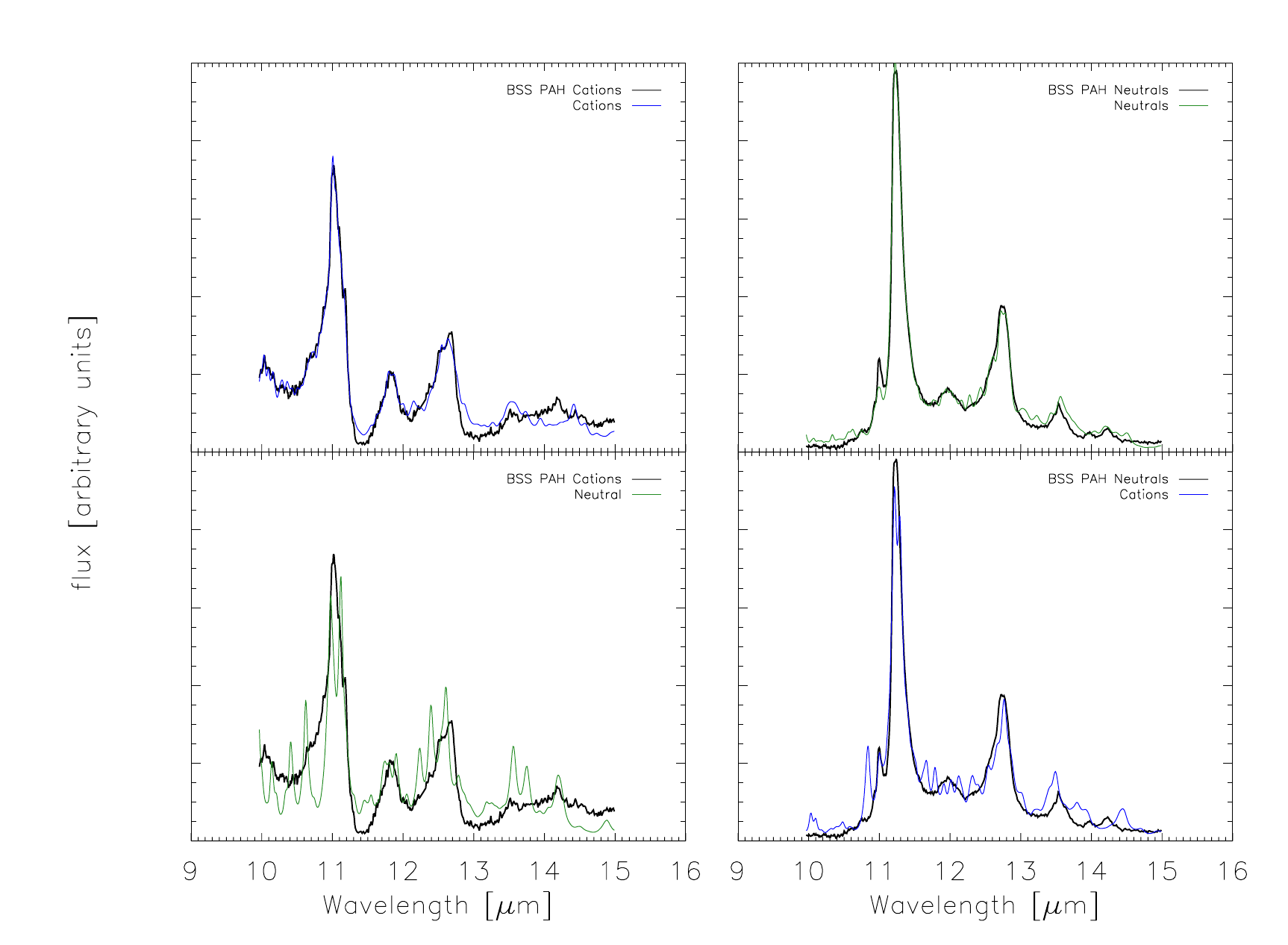}
\caption{PAH Database fit results of fitting the BSS extracted PAH$^+$ (left) and PAH$^0$ (right) signals using only cations (blue) or only neutral species (green).  The black line represents the original BSS extracted spectra while the fit is represented in color.}
\label{fig:degen}
\end{figure*}

To investigate the level of degeneracy of the database fit, we have performed several tests.  Here, we present four fits with explore this important issue.  All fits follow the same values described in Section~\ref{sec:dbfit}: a 6 cm$^{-1}$ FWHM, a 15 cm$^{-1}$ band redshift, a Lorentzian profile,  a 6 eV excitation energy with cascade emission, and molecules including only Carbon, Hydrogen, and Nitrogen.  The fits are described below.
\begin{enumerate}
\item The BSS extracted PAH$^+$ spectra is fit with the database restricting it to \emph{only} ionized species (top left of Figure~\ref{fig:degen}).
\item The BSS extracted PAH$^0$ spectra is fit with the database restricting it to \emph{only} neutral species (top right of Figure~\ref{fig:degen}).
\item The BSS extracted PAH$^+$ spectra is fit with the database restricting it to \emph{only} neutral species (bottom left of Figure~\ref{fig:degen}).
\item The BSS extracted PAH$^0$ spectra is fit with the database restricting it to \emph{only} ionized species (bottom right of Figure~\ref{fig:degen}).
\end{enumerate}
While the database fit of the BSS PAH$^+$ signal with PAH$^+$ cations is excellent, when it is limited to neutral species (bottom left Figure~\ref{fig:degen}), it is clear that the PAH$^+$ spectrum cannot be recreated.  However, the database fit of the BSS PAH$^0$ species with cations does provide a suitable fit, except for one key region: the 11.2 $\mu$m feature.  The best fit of the database produces a two-pronged feature with neither prong peaking exactly at the 11.2 $\mu$m band.  However, it is possible that by varying the FWHM and band shift, which are somewhat arbitrary values, these bands could blend to recreate the 11.2 $\mu$m feature.  This is due to the fact that various PAH cations peak throughout the mid-IR spectrum.  Specifically, PAH cations have emission features that peak throughout the 10.8 - 11.3 $\mu$m range.  This only highlights again the need for both approaches to apply as many astronomical constraints to the emitting species as possible, including wider spectral coverage when available.  The BSS method allows us to spatially separate the three emitting components and therefore we create a physical picture in which Signal 2, located nearest to the star, represented the PAH$^+$.  In this case, the database supports this claim since we cannot fit this spectra with only neutral PAH species.  We then move to Signal 1, which has a much different emission spectrum than the PAH$^+$ and is located in the middle of the PDR with a much higher abundance than the PAH$^+$.  The database shows that this can be fit with either only neutral or only cation PAH species, but the physical intuition of BSS indicates that this is most likely a neutral PAH spectrum.

Essentially, the database is a collection of Lorentzians that when restricted to a narrow band pass in which aromatic bands fall, and added together with varying intensity, can likely fit any type of aromatic-rich materials.  Especially in the case of cations, the fit is degenerate (Figure~\ref{fig:degen}).  Yet when the fit is performed with the full range of the database, the highest $\chi^2$ solution is strikingly similar to the BSS extracted solution (Figure~\ref{fig:fitvbss}).  Although our tests show that the fits are poorly constrained for individual PAHs, this is not the case when considering PAH sub populations, which trace properties such as size
and charge.  Thus, the astronomical mid-IR bands studied here are assigned to PAH molecules as a class, not to individual PAH molecules. The literature is rich in examples where it is shown that the mid-IR PAH emission features can be categorized by charge, size, shape and composition \citep{peeters2,baus,baus09,2010A&A...511A..32B}. When a single species is removed from the pool of species that can be fit, another species from the same subgroup will simply replace it. Therefore, a mid-IR fit made with the database is expected not to provide information about the presence of an individual PAH molecule, but about the subclasses of PAHs involved and the response of that subclass to the local astronomical environment. 

\begin{figure*}
\centering
\includegraphics[width=17cm]{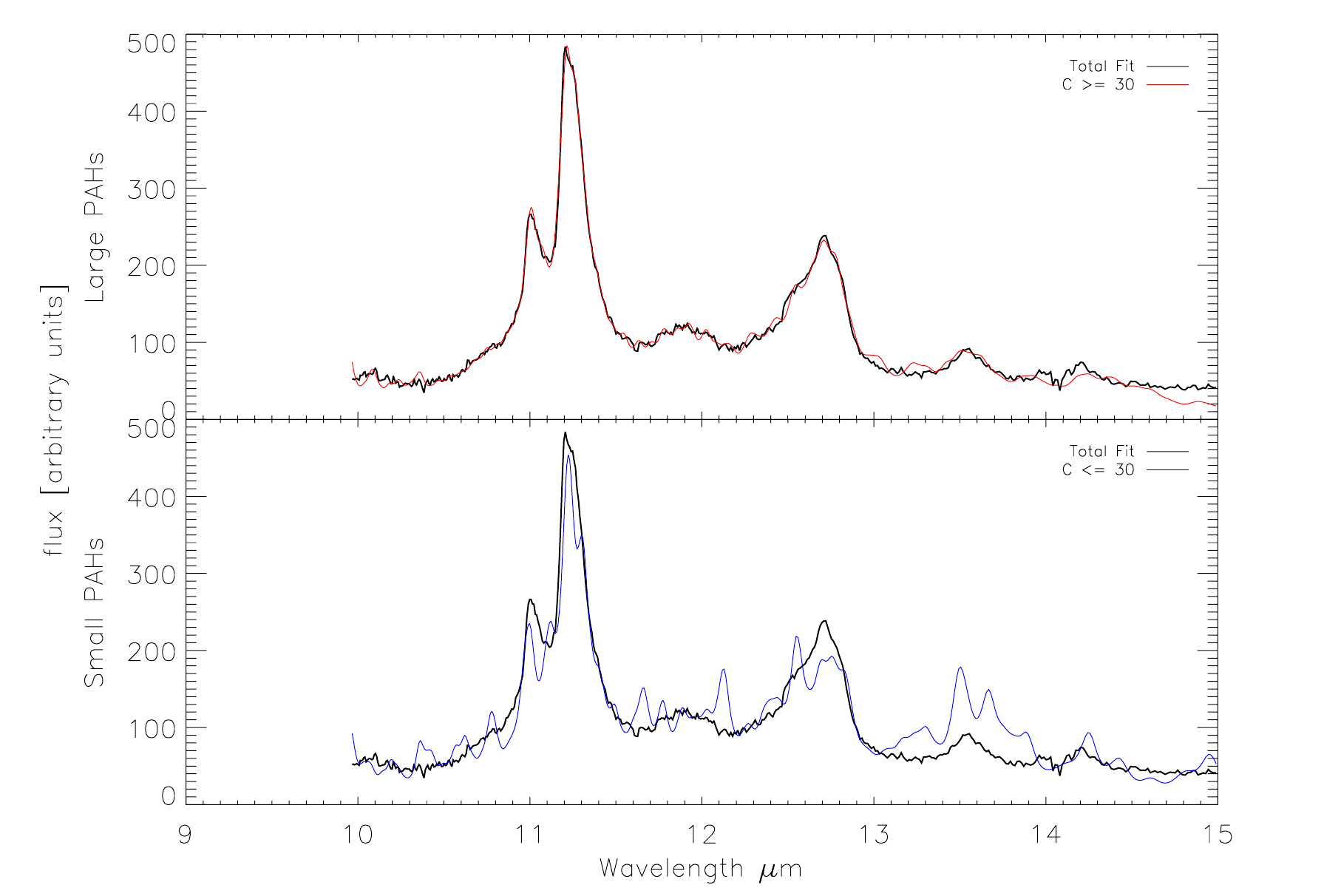}
\caption{PAH Database fit results of fitting an observed spectrum limiting the species to large PAHs (C $\ge 30$) (top) and small PAHs (C $\le 30$) (bottom). The black line represents the original observed spectrum, where the fit is represented in color.}
\label{fig:size}
\end{figure*}

Aside from charge, the database also probes the subclass of size.  Although we have a limited wavelength range, the main features are best reproduced by larger PAH molecules.  Similarly to the case of charge, we can treat PAH size as another sub population with a set of characteristic properties.  In Figure~\ref{fig:size}, we use the database to fit the observed spectra from Figure~\ref{fig:fitvbss}.  Here, besides size the only constraint applied is including only C, H, and N atoms.  The parameters are equal to those described in Section 5.1.  The top panel displays a fit of the observed spectrum using only PAHs with 30 or more C atoms.  The bottom panel is a fit of the observed spectrum using PAHs with 30 or less C atoms.  There are 255 PAH species available to fit C $\le 30$, whereas there are 237 PAH species available to fit C $\ge 30$, once the other constraints are taken into account.  Even though there are less PAH species included in the fit of large PAHs, it is a superior fit to the small PAHs and has a lower reduced norm value.  From this limited wavelength range, we can already deduce that large PAHs are responsible for the main bands, however to have a stronger constraint on size, an analysis using the full 3 - 15 $\mu$m range in necessary.  

\bibliographystyle{aa}
\bibliography{bib_file.bib}
\listofobjects
\end{document}